\documentclass[proof]{gen-bioinformatics}
\usepackage{graphicx}

\makeatletter
\renewcommand\boldmath{\@nomath\boldmath\mathversion{bold}}
\makeatother

\usepackage{amssymb,amsfonts,url,times}
\usepackage{hlight}
\def\email#1{#1}

\begin{document}
\Sup{1}

\verso{T. P\"oschel and J.A. Freund}
\recto{How to decide whether small samples comply with an equidistribution}

\title{How to decide whether small samples comply with an equidistribution}

\author[1]{\pfnm{Thorsten}
 \pinit{}
 \psnm{P\"oschel}}

\author[2]{\pfnm{Jan}
 \pinit{A}
 \psnm{Freund}}

\address[1]{\porgdiv{Institut f\"ur Biochemie}
 \porgname{Charit\'e}
 \pstreet{Monbijoustra{\ss}e 2}
 \pcity{Berlin}
 \postcode{D-10117}
 \pcnty{Germany}}

\address[2]{\porgdiv{Institut f\"ur Physik}
 \porgname{Humboldt-Universit\"at zu Berlin}
 \pstreet{Invalidenstra{\ss}e 110}
 \pcity{Berlin}
 \postcode{D-10115}
 \pcnty{Germany}}

\maketitle

\begin{abstract}
\noindent
{\bf Motivation:} The decision whether a measured distribution
complies with an equidistribution is a central element of many
biostatistical methods. High throughput differential expression
measurements, for instance, necessitate to judge possible
over-representation of genes. The reliability of this judgement,
however, is strongly affected when rarely expressed genes are pooled.
We propose a method that can be applied to frequency ranked
distributions and that yields a simple but efficient criterion to
assess the hypothesis of equiprobable expression levels.
\\
{\bf Results:} By applying our technique to surrogate data we exemplify
how the decision criterion can differentiate between a true
equidistribution and a triangular distribution. The distinction
succeeds even for small sample sizes where standard tests of
significance (e.g. $\chi^2$) fail. Our method will have a major impact
on several problems of computational biology where rare events baffle a
reliable assessment of frequency distributions.
\\
{\bf Availability:} The program package is available upon request from
the authors.
\begin{subabstract}[Contact]
\email{thorsten.poeschel@charite.de}
\end{subabstract}
\end{abstract}
\baselineskip12pt plus .2pt minus .05pt

\section{Introduction}

Biostatistical analyses quite generally infer desired information from
experimentally observed frequency distributions. Microarrays, for
instance, can screen thousands of genes simultaneously, thus allowing
for high-throughput measurements \cite[]{microarray}. The expression
level of genes is quantified via hybridisation signal intensities and,
after an appropriate normalisation procedure, yields a vector of
numbers which can be understood as an empirically obtained frequency
distribution. The important issues of coregulation \cite[]{coreg} and
differential expression \cite[]{diffexp} are based on further analysis
of these distributions. In this context, pooling a family of rarely
expressed genes can be combined with the task to decide whether the
observed expression pattern complies with an equidistribution
\cite[]{p53}. As is intuitively clear, the small sample size makes this
decision a hard problem - nobody would expect to reproduce the
statistics of a fair die (with equal probability $1/6$ for each side)
with only four trials. The problem of small sample statistics pervades
literature, e.g.~\cite[]{ssp1,ssp2,ssp3}.  Similar problems might occur
in the statistical analyses of codon distributions
\cite[]{codon} or other biopolymers \cite[]{protexpr}.  As a
last field of potential application let us mention the computational
comparison of two draft sequences of the human genome
\cite[]{draftseqs}.

We propose a method to decide the above posed question of an
equidistribution based on frequency ranked statistics. The technique
yields a criterion that can detect frequency distributions generated
from a true equidistribution and reject others. It is important to
note that the criterion we have devised is rather efficient for small
sample sizes (expression levels) where standard measures of
significance like the $\chi^2$-test or the Kolmogorov-Smirnov test fail. We expect that
biostatistical analyses of small sample data, as met e.g.~for rarely
expressed genes, can profit from our criterion.

\section{System and methods}

Assume, in an experimental sampling probe we find $N^*$ different
species occurring with relative frequencies $f_1$, $f_2$ \dots
$f_{N^*}$. The term {\em species} is not meant in its strict biological sense here but more general as a class of individuals. As an example we mention the subsequences of a certain length, e.g. $l=6$, of the nuclein acids $A$, $G$, $C$, and $T$. A species is the word $GATAGG$ which may be found in a gene at various positions, e.g., at positions 45, 122, and 431. For this example we say that the species $GATAGG$ occurs with 3 representatives.  

Quite often uniform probabilities are motivated by
theoretical considerations or as the simplest assumption. We present a
practical criterion based on finite sample size statistics that allows
to decide, i.e.~to accept or reject, the hypothesis of an underlying
uniform probability distribution $p_1=p_2=\dots=p_N=1/N$ with $N\ge
N^*$.

Given a probability distribution $\{p_i,~i=1,\dots,N\}$ with
$\sum_{i=1}^N p_i=1$  where the $p_i$ represent the likelihood to find a
representative of species $i$ from a set of $N$ possible
species.  Assume, in an experiment we find $M_{1}$ copies of
species $1$, $M_{2}$ copies of species $2$ \dots $M_{N^*}$
copies of species $N^*$, with $\sum_{i=1}^{N^*} M_i=M$ and none of
the remaining species $N^*+1,\ldots, N$. The fact that not all of
the $N$ species might be observed means ~$N^*\le N$. Stated
differently, the number $N$ is not directly
accessible by the experiment but has to be estimated using
probabilistic arguments. However, beyond all methods of optimally
estimating the unknown $N$ -- which then completely specifies the
assumed equidistribution -- the question arises how probable the
hypothesis of an equidistribution is at all.

Clearly, the law of large numbers asserts the stochastic convergence
of relative frequencies towards the related probabilities, i.e.
\begin{equation}
\label{eq:pdef}
p_i\equiv\lim\limits_{M\rightarrow\infty} f_i\,,~~~~~f_i\equiv\frac{M_i}{M}\,,
\end{equation}
where the limit is approached {\em for almost all} (in the mathematical sense) experimental
realizations (sampling probes). For cases where the condition $M\gg N$ is not fulfilled, however, the
relative frequencies often deviate considerably from the related
probabilities, e.g. \cite{Schmitt,Herzel,PER}. Strictly speaking, for
finite $M$ one cannot even be sure to have found each species at least
once. Just imagine one or a few species with probabilities being
orders of magnitude smaller than the overwhelming rest of nearly
identically probable species. One might say that in such a situation
all the tiny probability events are dispensable for an efficient
description and a uniform distribution is true rather in a practical
sense. However, in other situations deviations of an underlying
probability distribution from a true equidistribution can be that
significant that the hypothesis of a uniform probability distribution
should be judged as inappropriate. As an example consider a triangle
shaped distribution. Now, how does this substantial discrepancy show
up in experimental sampling probes and how can this be distinguished
from pure finite sample size effects?

\begin{figure}[htbp]
  \resizebox{\columnwidth}{!}{
\includegraphics{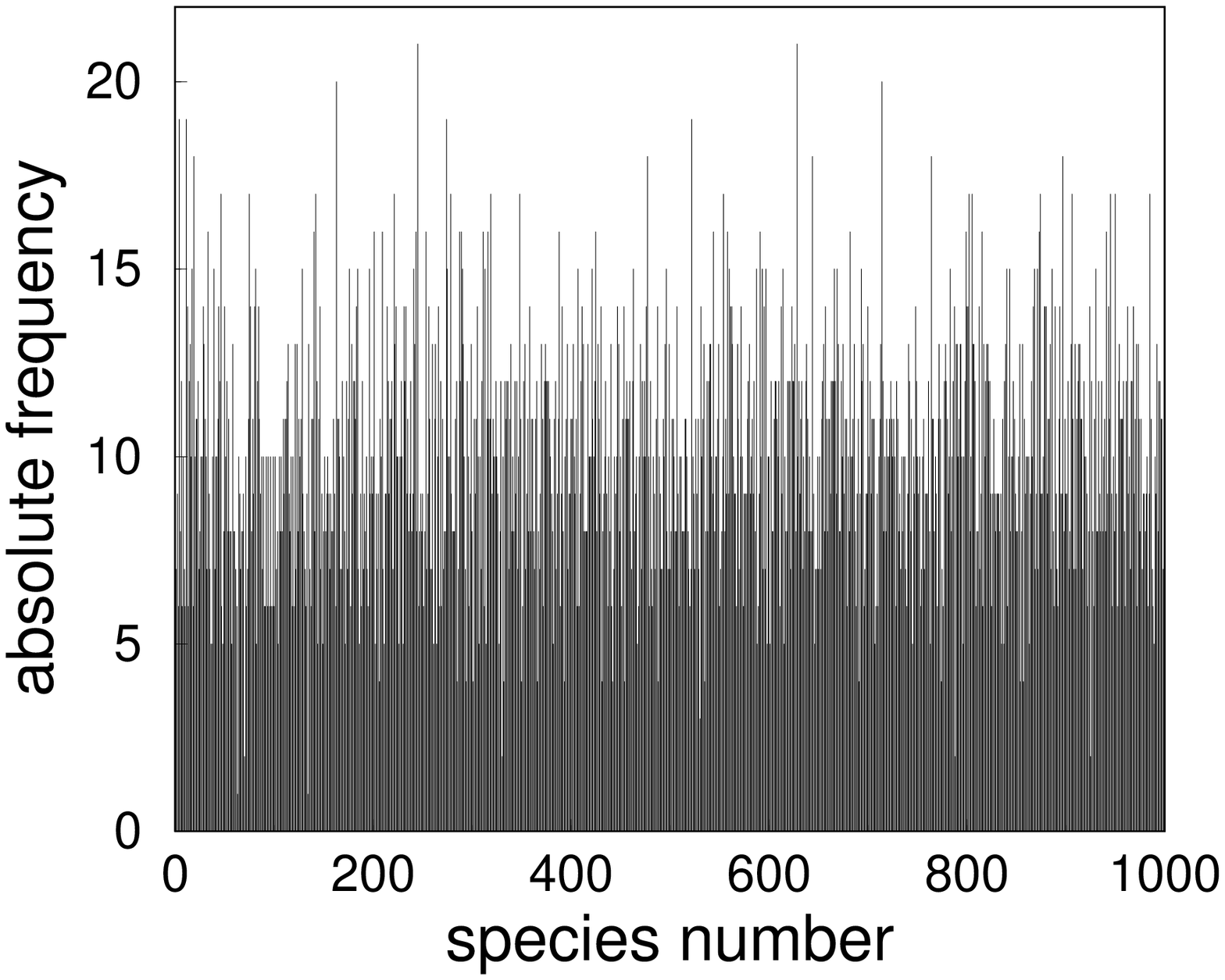}
\includegraphics{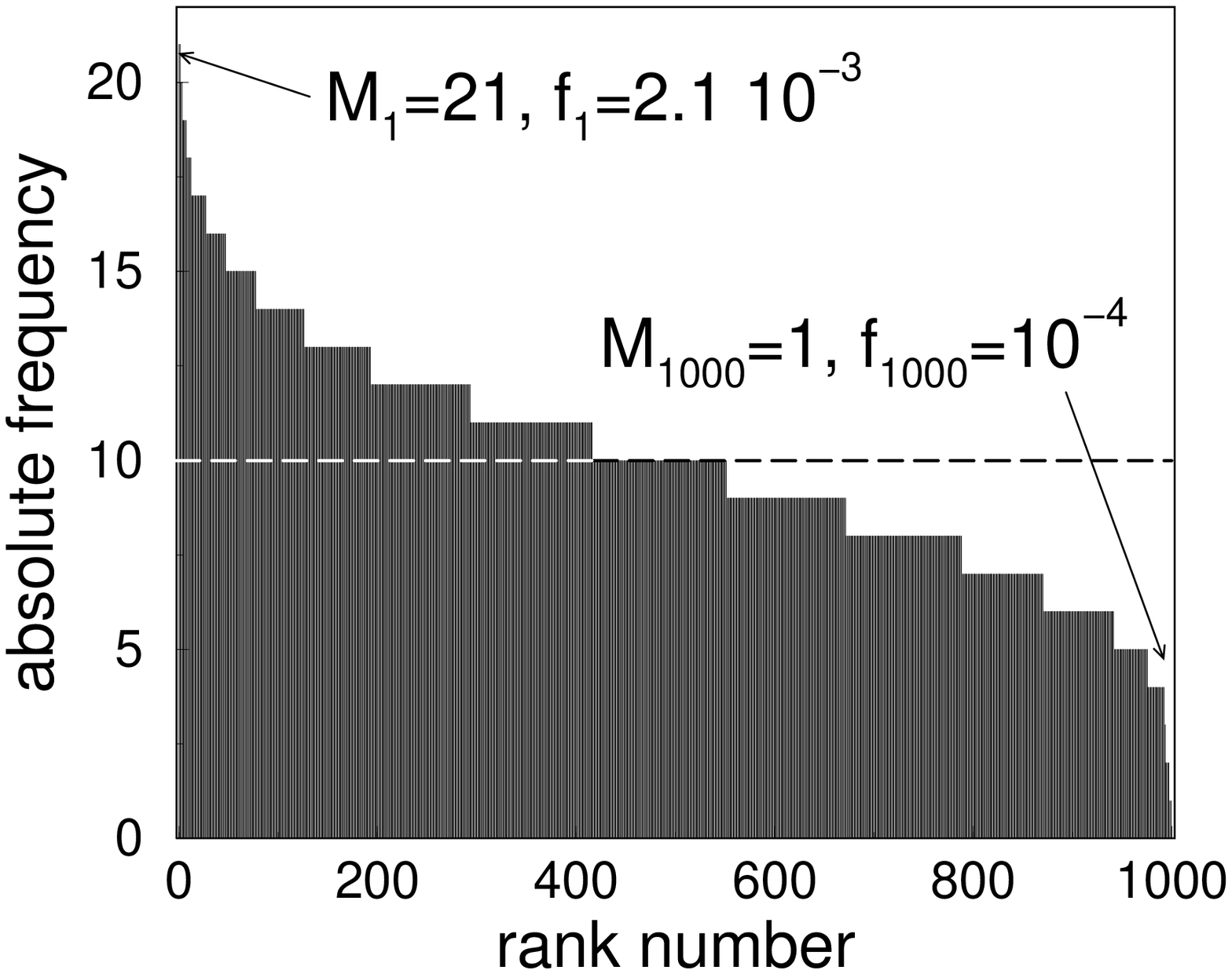}
  }   
\vspace*{-0.3cm}
  \caption{The measured frequency distribution may deviate from the
  probability distribution considerably due to finite sample
  effects. The left figure shows absolute frequencies which have been
  obtained by drawing $M=10^4$ random integers from the interval
  $1\dots N$, $N=1000$. The right figure presents the same data but in
  rank order (see text). Whereas the expectation value for each of the
  numbers is $M/N=10$ (dashed line) large deviations are noticeable.}
  \label{fig:zipf}
\end{figure}

To illustrate typical distortions of the equidistribution due to finite
sample size we depict Fig. \ref{fig:zipf} (left) which shows a
frequency distribution obtained by drawing $M=10^4$
equidistributed random integers from the interval $[1,2,\dots
1000]$, i.e., $N=1000$. From the probabilities $p_1=p_2=\dots p_N=10^{-3}$ we expect
to find each of the numbers, on average, $M/N$ times, i.e.,
$\left<f_1\right>=\left<f_2\right>= \dots = \left<f_N\right>=
10$. Figure \ref{fig:zipf} shows that there are large fluctuations of
the occurrences of the numbers. A more convenient way to represent
these data is the rank ordered frequency distribution. To this end we
re-order the abscissa in a way to receive a decaying curve of
frequencies, i.e., the most frequent species occupies rank 1, the most
frequent but one occupies rank 2 etc. The rank ordered frequencies are
depicted in the right part of Fig. \ref{fig:zipf}.

Since our random numbers are equidistributed by construction the
expectation value for each of the numbers (ranks) is $M/N=10$. As seen
from Fig.~\ref{fig:zipf} fluctuations around this mean are
considerable: if one naively inferred the probabilities (or
concentrations) from the relative frequencies, one would end up with
a relative error of 110\% for small ranks and 90\% for large ranks. Being
faced with a measurement as the one sketched in Fig.~\ref{fig:zipf} it
is far from trivial to decide whether the objects (in our case random
numbers) are equidistributed. It is the aim of this paper to
propose a method which allows to distinguish between uniform and
non-uniform probability distributions if the sample size is too small
to identify the probabilities with the relative frequencies due to
Eq. (\ref{eq:pdef}).

\subsection{Finite size statistics}
For equidistributed events $j\in [1\dots N]$ with $p_j=p=1/N$ the
probability to find with $M$ trials a number of {\em at least} $k_i$
different events each occurring exactly $i$ times with $i=0,\dots,M$
reads \cite[]{vonMises}
\begin{equation}
  \label{eq:atleast} P\!\left(k_i,i\right)\! =\! \frac{M!}{(i!)^{k_i}
  \!\left(M\!-\!k_i i\right)!} \left(\!\frac 1 N\!\right)^{\!k_i
  i}\!\left(1\!-\!\frac{k_i}{N}\right)^{\!\left(M-k_i i\right)}\,.
\end{equation}
Applying the Exclusion-Inclusion-Principle \cite[]{JK} to
Eq. (\ref{eq:atleast}) one can derive the probability to find with $M$
trials {\em exactly} $k_i$ different species each occurring exactly $i$
times:
\begin{eqnarray}
&& \!\!\!\!\!\!\!\!\!\! p(k_i,i)=\sum\limits_{j=k_i}^{\lfloor
M/i\rfloor}(-1)^{(j-k_i)} \left(j\atop k_i\right)P(j,i)\nonumber \\
&&\!\!\!\!\!=\frac{M!}{N^M}\sum\limits_{j=k_i}^{\lfloor M/i\rfloor}
(-1)^{(j-k_i)}\left(j\atop k_i\right) \frac{(N-j)^{(M-ji)}}{(i!)^{j}
(M-ji)!}\,,  \label{eq:InEx}
\end{eqnarray}
where $\lfloor x \rfloor$ stands for the integer of $x$.
The first moment of this probability reads
\begin{equation}
  \left<K_i\right>= \left(M\atop i\right) N^{(1-i)}  \left(1-\frac{1}{N}\right)^{(M-i)}.
\label{eq:Mom}
\end{equation}
Hence, when randomly drawing $M$ representatives each occurring with
the same probability $p_1=p_2=\dots p_N$, one expects to find
$\left<K_0\right>$ species zero-times, $\left<K_1\right>$ once, \dots
$\left<K_M\right>$ species $M$ times. The full derivation of
Equations (\ref{eq:InEx}) and (\ref{eq:Mom}), using rather involved
algebra, can be found in \cite{JT,TJ}.

For our above example, if $M$ random integers have been drawn from the
interval $1,\dots,N$, Eq. (\ref{eq:Mom}) describes how many random
numbers, on average, occur exactly zero times, i.e., never
($\left<K_0\right>$), once ($\left<K_1\right>$), twice
($\left<K_2\right>$), \dots, $M$ times ($\left<K_M\right>$). Hence,
using Eq. (\ref{eq:Mom}), for an equidistribution it is possible to
calculate the expected measured frequencies \cite[]{TPJanf}.

In converse direction, it is possible to infer for each $i=1\dots M$ a value 
$N^{(i)}$ from the experimentally observed $k_i$ by identifying
$k_i=\left<K_i\right>$ ($i=1\dots M$) and making use of
Eq. (\ref{eq:Mom}). Thus, sampling a true equidistribution, one should
find
\begin{equation}
  \label{eq:N} 
  N^{(1)}\approx N^{(2)}\approx \dots\approx
N^{(L)}\approx N \,.
\end{equation}
In theory, the distribution (\ref{eq:InEx}) exists for $i=1\dots M$ where the event related to $k_M$ corresponds to the extreme case that all $M$ representatives belong to the same species. In measurements not all $k_i$ can be different from zero. Therefore, the approximative equation (\ref{eq:N}) is valid for all upper indices $(L)$ for which the corresponding $k_L\ne 0$ has been found in the measurement.

The estimated values read for $i=1$:
\begin{equation}
  \label{eq:N1}
  N^{(1)}=\left[1-\left(\frac{\left<K_1\right>}{M}\right)^\frac{1}{M-1}\right]^{-1}\,,
\end{equation}
and for all other occupation numbers $\left<K_i\right>$
\begin{equation}
  \label{eq:Ni} N^{(i)}=\left[\left(M\atop i\right)
  \frac{1}{\left<K_i\right>}\left(1-\frac 1
  {N^{(i)}}\right)^{M-i}\,\right]^{i-1}\,.
\end{equation}
Equation (\ref{eq:Ni}) has to be solved numerically by an iteration
procedure. As discussed below
in dependence
on the variables $M$, $\left<K_i\right>$, and $i$ this equation may
have zero, one, or two solutions and one has to select the
appropriate one.

To clarify the meaning of the approximate identity signs in
Eq.~(\ref{eq:N}) we point out that the identification
$k_i=\left<K_i\right>$ is an approximation which should be amended by
statistical fluctuations, i.e., $k_i=\left<K_i\right>+\left<\Delta K_i\right>$, with
$\left<\Delta K_i\right> \sim \sqrt{\mbox{var}K_i}$. The variance of $K_i$
for an equidistribution can be achieved analytically using the generating function of $p\left(k_i,i\right)$ (see Eq. (\ref{eq:InEx})). It reads \cite[]{TPJanf}
\begin{eqnarray}
   \label{var2}
   \mbox{var}K_i &=&\left<K_i^2\right>-\left<K_i\right>^2\\
        &&\!\!\!\!\!\!\!\!\!\!\!\!\! = \left<K_i\right>\! 
        \left[1\!+\!\left(\!M\!-\!i\atop i\!\right){(N-2)^{M-2i}\over
        (N\!-\!1)^{M-i-1}}\right] \!- \left<K_i\right>^2\nonumber\,.
\end{eqnarray}
This variance of the $K_i$ can be converted into a characteristic error interval
around the derived $N_i$ simply by applying Eq. (\ref{eq:Mom}) not
only to $K_i$ but also to $K_i+\Delta K_i$ and $K_i-\Delta K_i$.

This means, if for each $i$ we plot the set of experimentally
determined numbers $N^{(i)}$ together with their expected range of
fluctuations ($\pm\sqrt{\mbox{var}N_i})$), for an underlying
equidistribution we should be able to draw a straight line which passes
through all the error intervals. On the contrary, if any horizontal
line significantly falls outside at least one of the intervals the
hypothesis of an underlying equidistribution should be rejected.

\section{Results}

We want to illustrate the method by means of an equidistribution
\begin{equation}
  \label{eq:pequi}
  p_i = 1/100\,,~~~~~~~ i=1,\dots,100
\end{equation}
and a triangular distribution
\begin{equation}
  \label{eq:ptriang}
  p_i=\frac{2}{100} \left(1-\frac{i}{101}\right)\,,~~~~~~~ i=1,\dots,100\,.
\end{equation}

\begin{figure}[htbp]
  \resizebox{\columnwidth}{!}{
\includegraphics{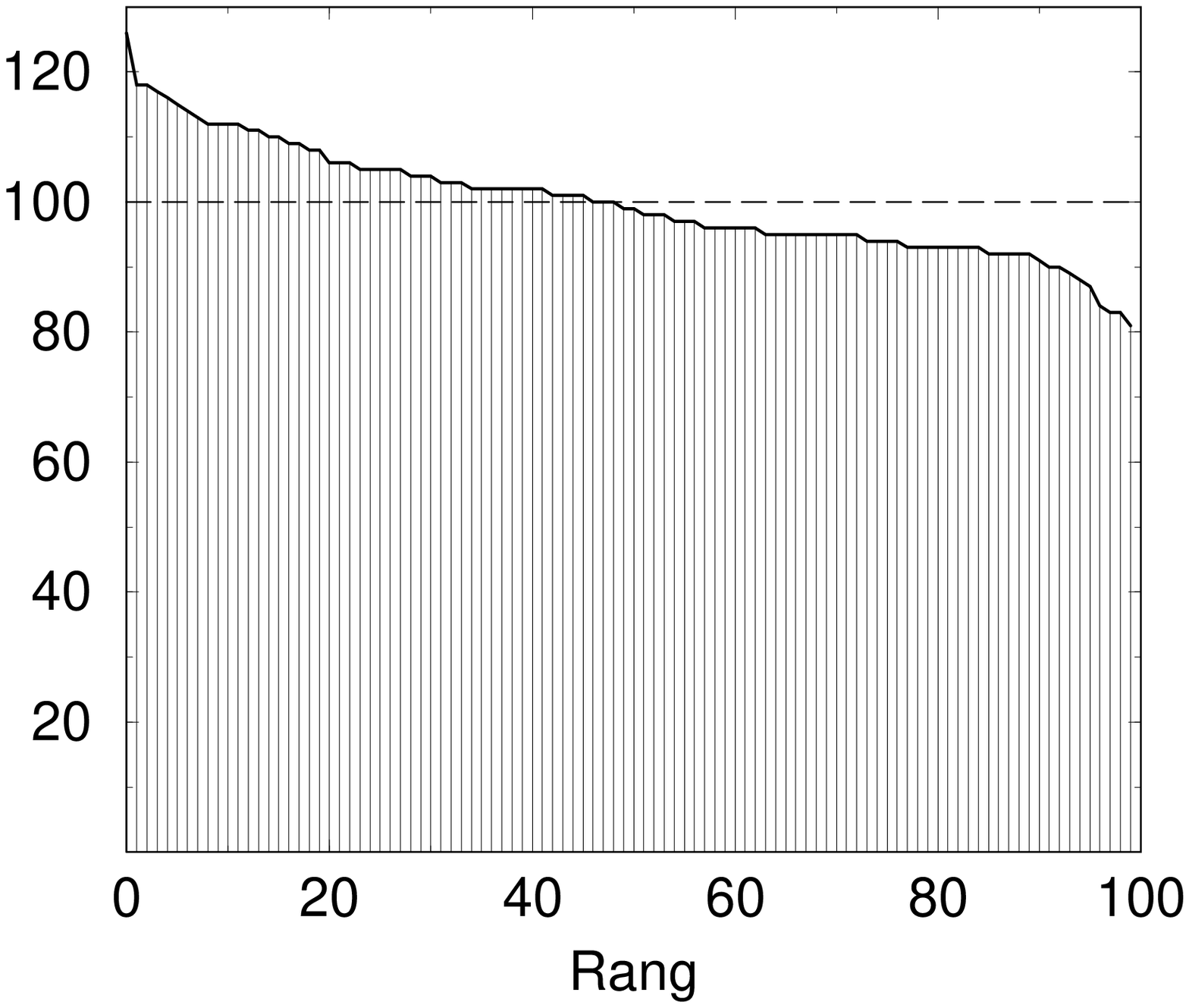}
\includegraphics{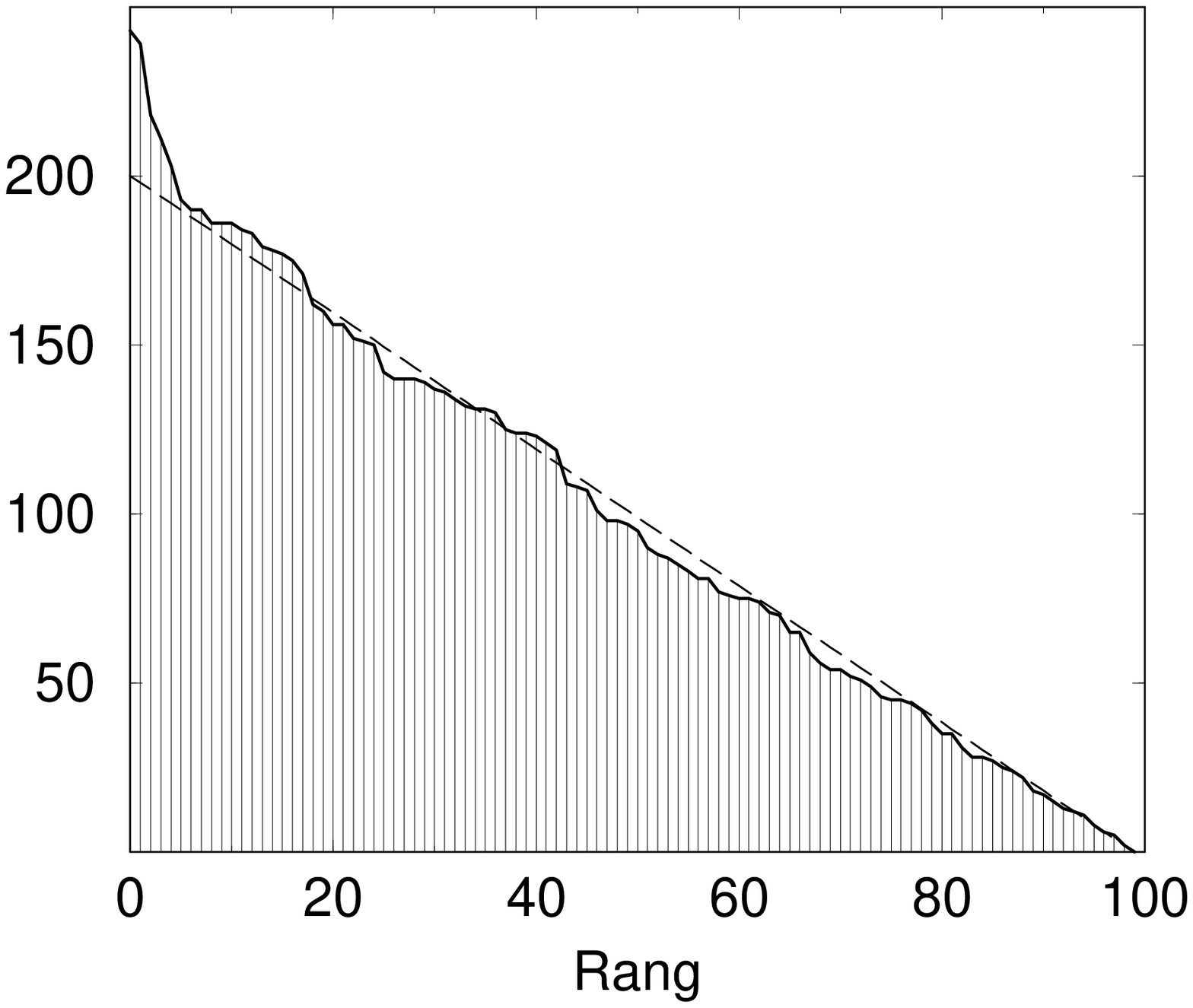}
  }   
  \resizebox{\columnwidth}{!}{
\includegraphics{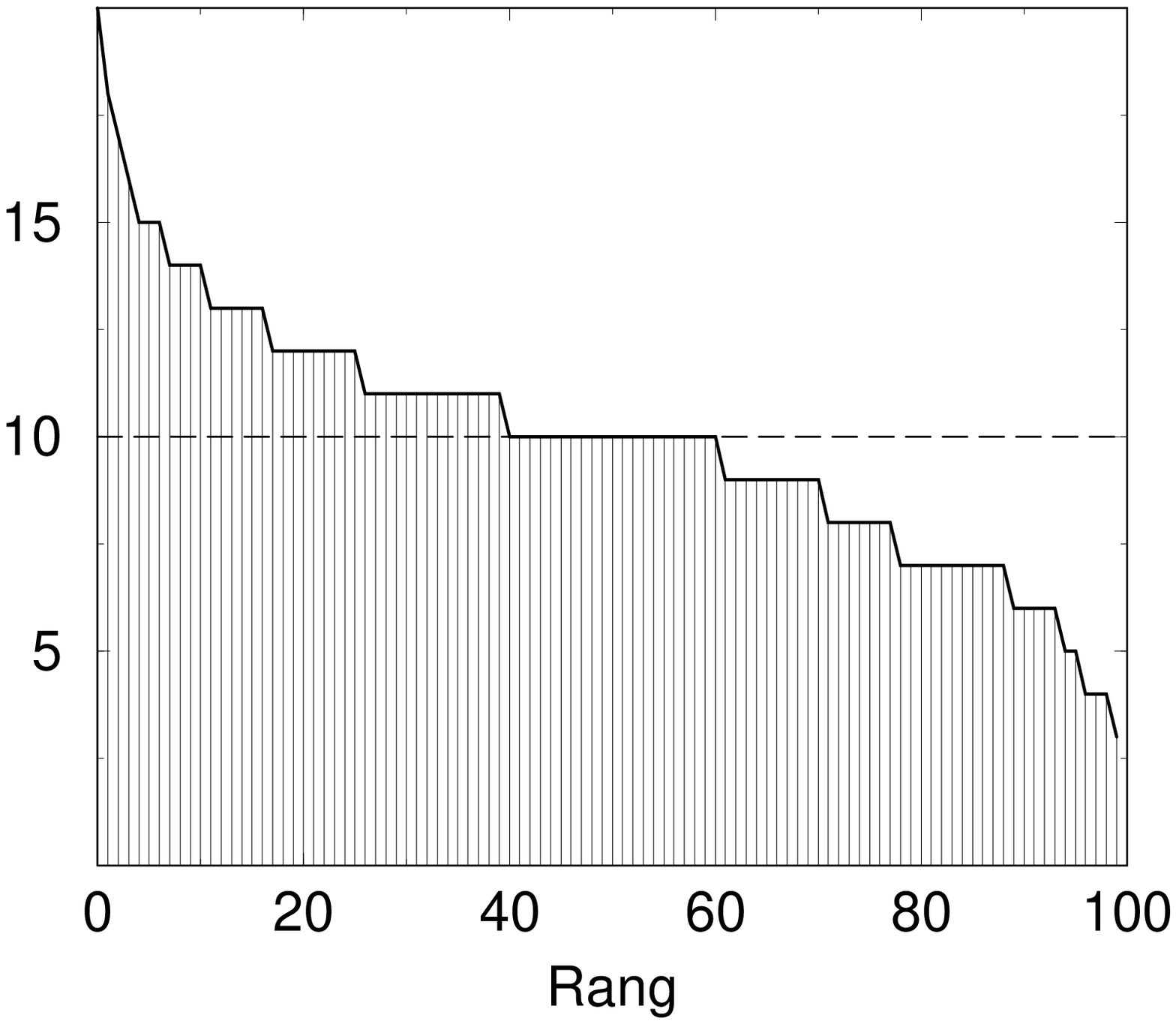}
\includegraphics{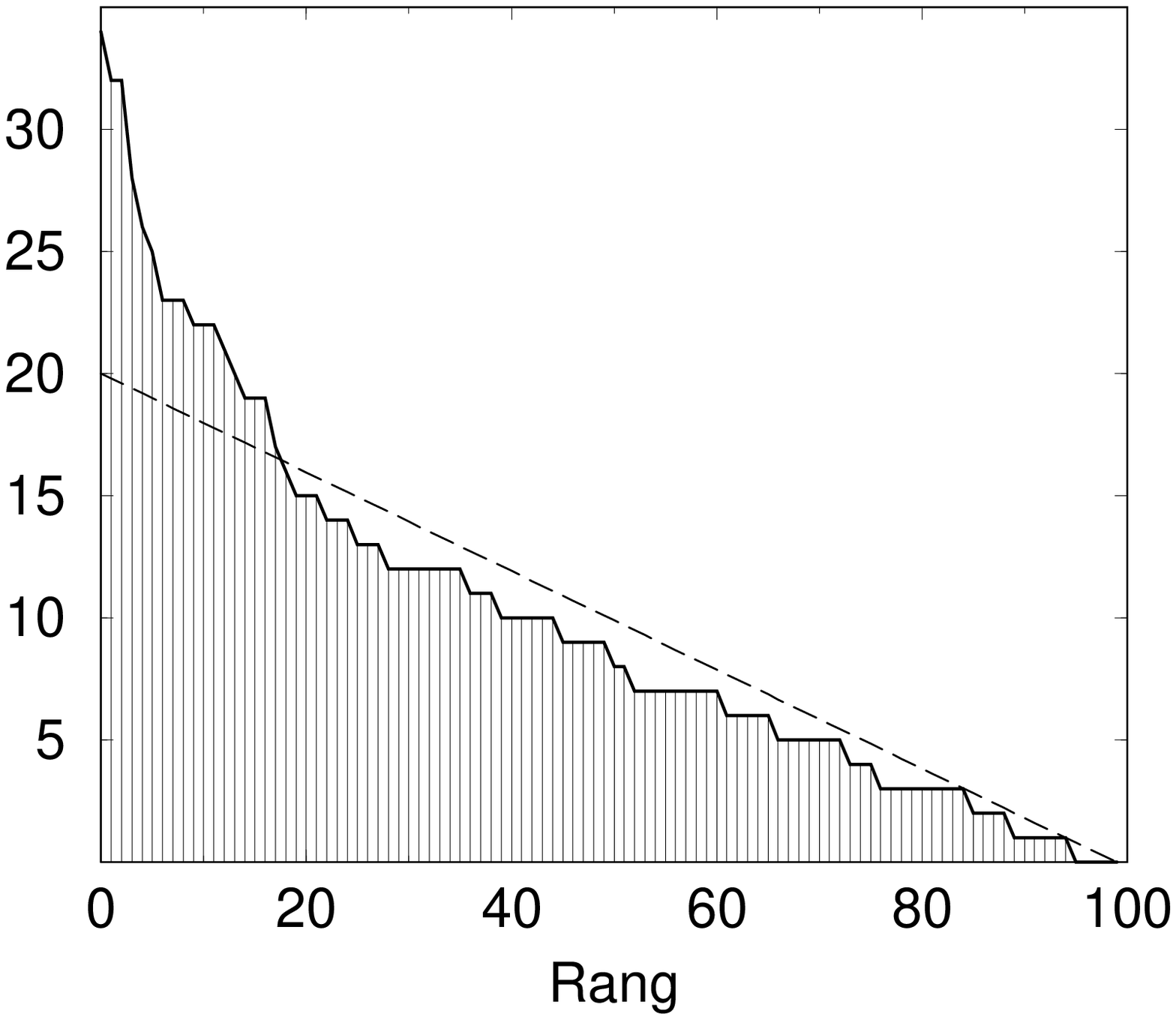}
  }   
  \resizebox{\columnwidth}{!}{
\includegraphics{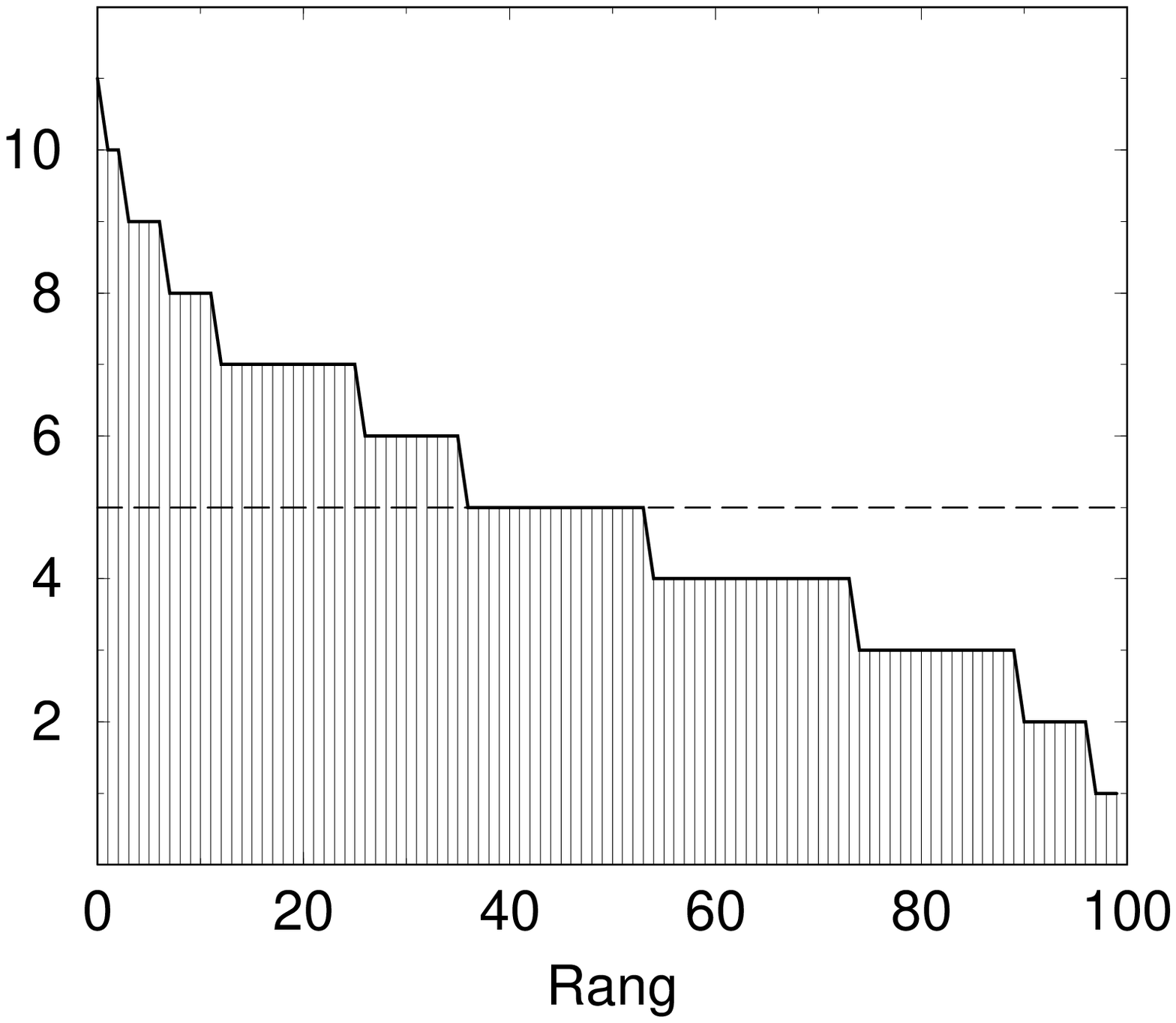}
\includegraphics{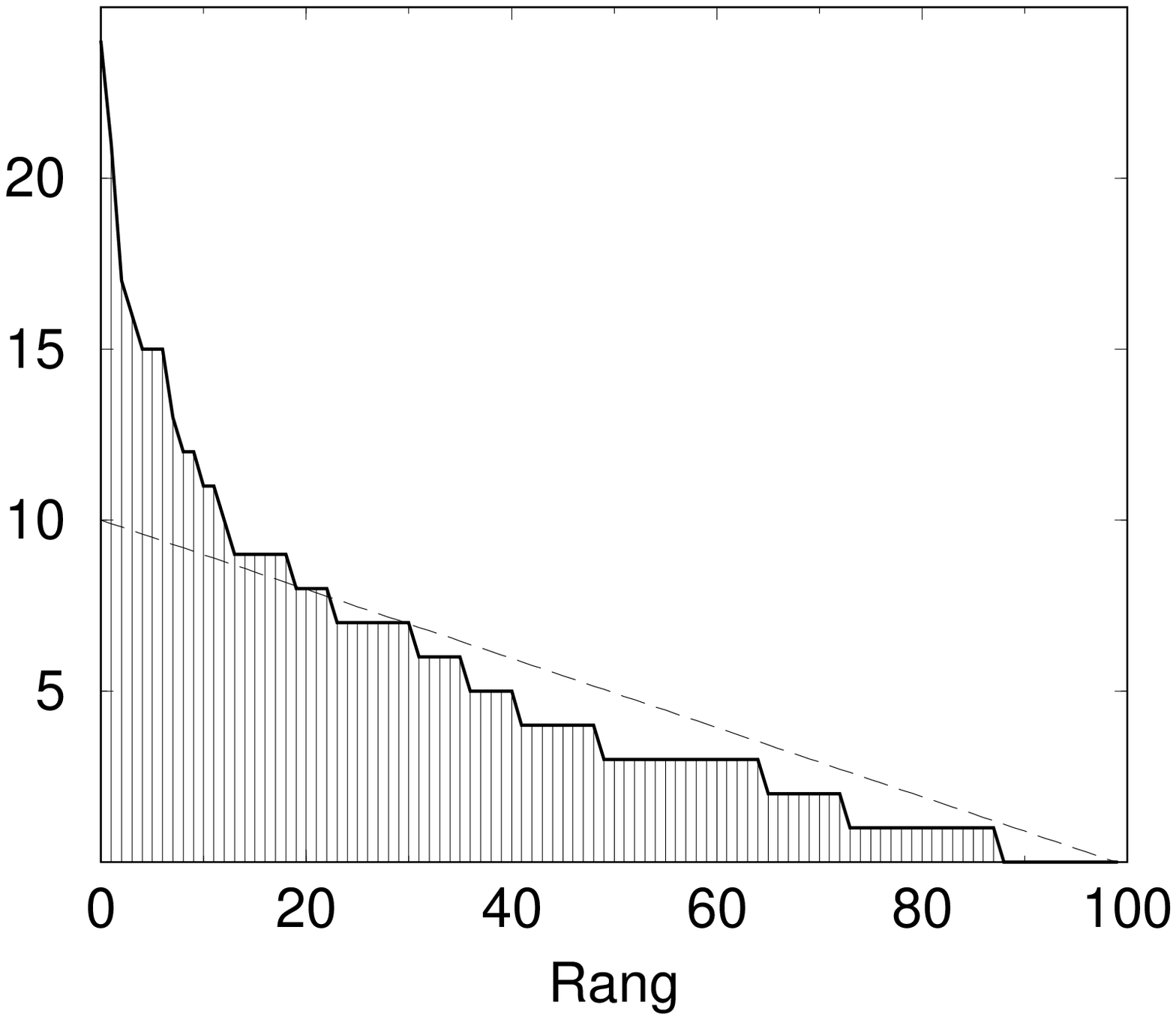}
  }   
  \resizebox{\columnwidth}{!}{
\includegraphics{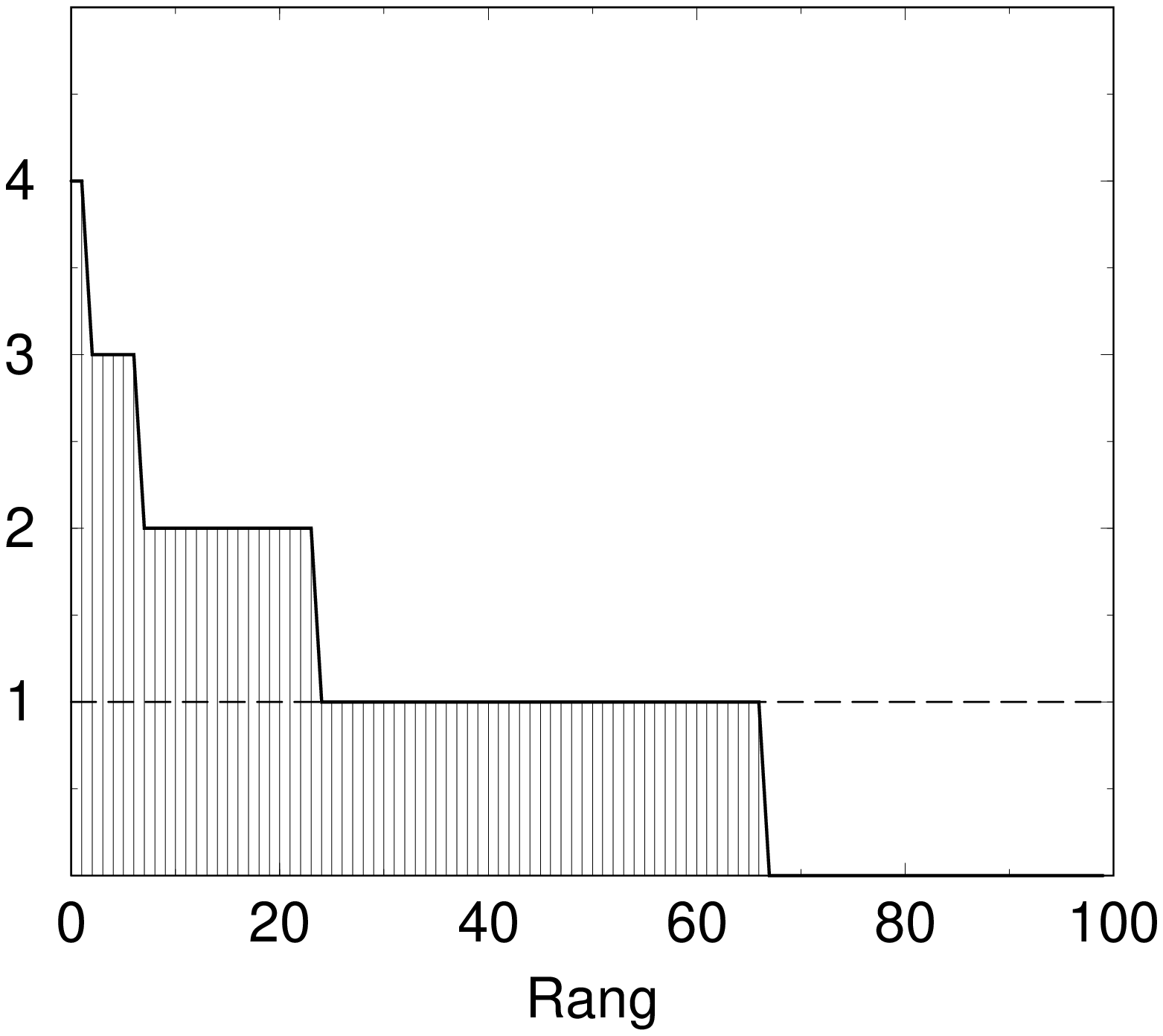}
\includegraphics{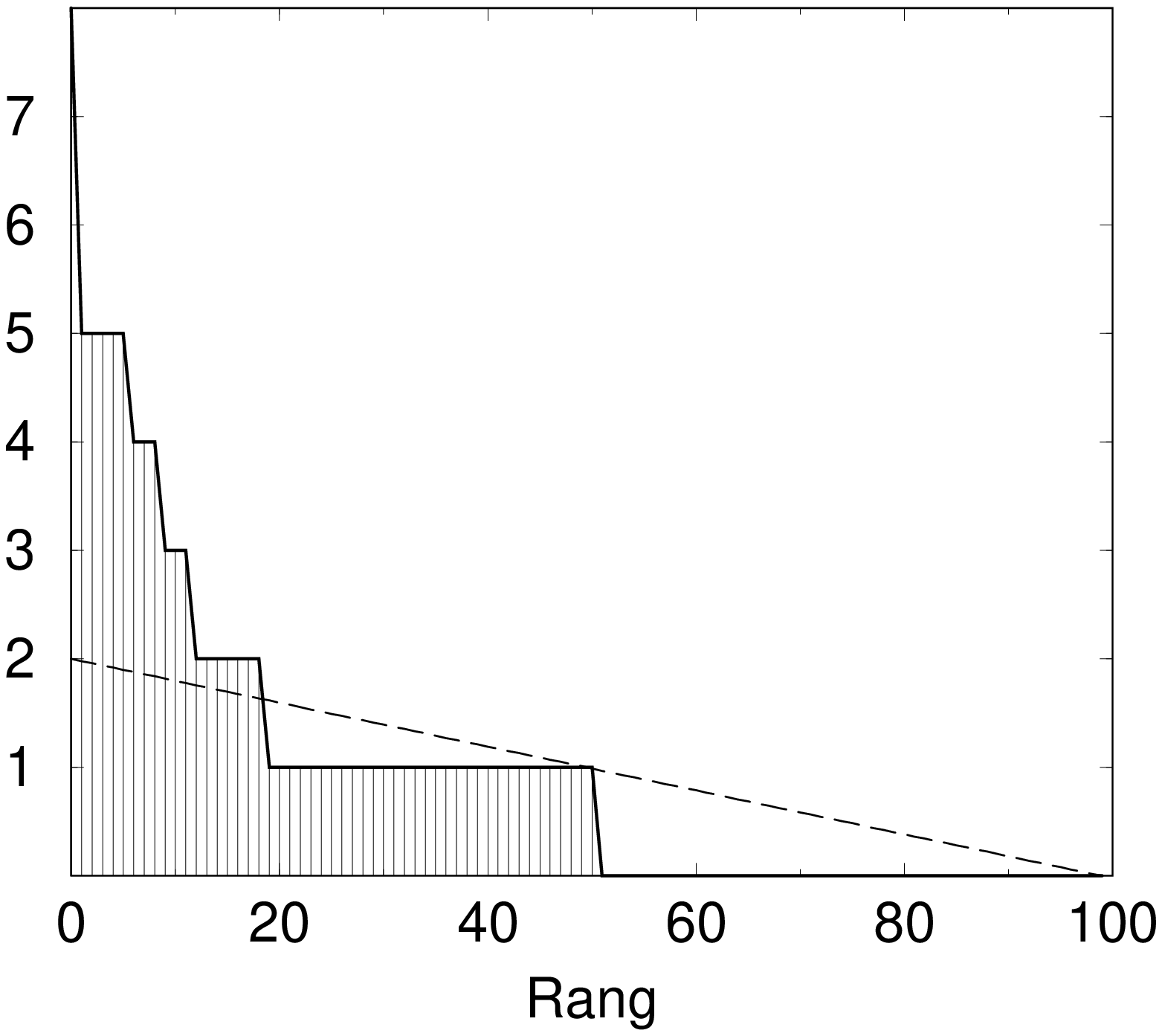}
  }   
\vspace*{-0.3cm}
\caption{Frequencies generated from the equidistributed probabilities, Eq. (\ref{eq:pequi}) (left plots) and from the triangular probability distribution Eq. (\ref{eq:ptriang}) for different sample sizes. From top to bottom: $M=10^4$, $M=10^3$, $M=500$, $M=100$. The dashed lines show the underlying probability distributions, Eq. (\ref{eq:pequi}) (left figures) and Eq. (\ref{eq:ptriang}), respectively.}
  \label{fig:deform}
\end{figure}

As discussed above, when drawing random events according to these
probability distributions, the resulting rank-ordered frequency
distributions will significantly depend on the sample size $M$. Figure
\ref{fig:deform} shows the rank-ordered frequency distributions for
different values of the sample size $M$. Whereas the top row figures
($M=10^4$) clearly reflect the equidistribution (left) and the
triangular distribution (right), respectively, for smaller sample sizes
(lower rows) one notices, as expected, significant deviations between
probabilities and frequencies. The central issue is now whether from
these small sample rank-ordered frequency distributions (lower rows)
one can nevertheless distinguish between the uniform and the
triangular probability distribution.

\begin{figure}[htbp]
  \resizebox{\columnwidth}{!}{
\includegraphics{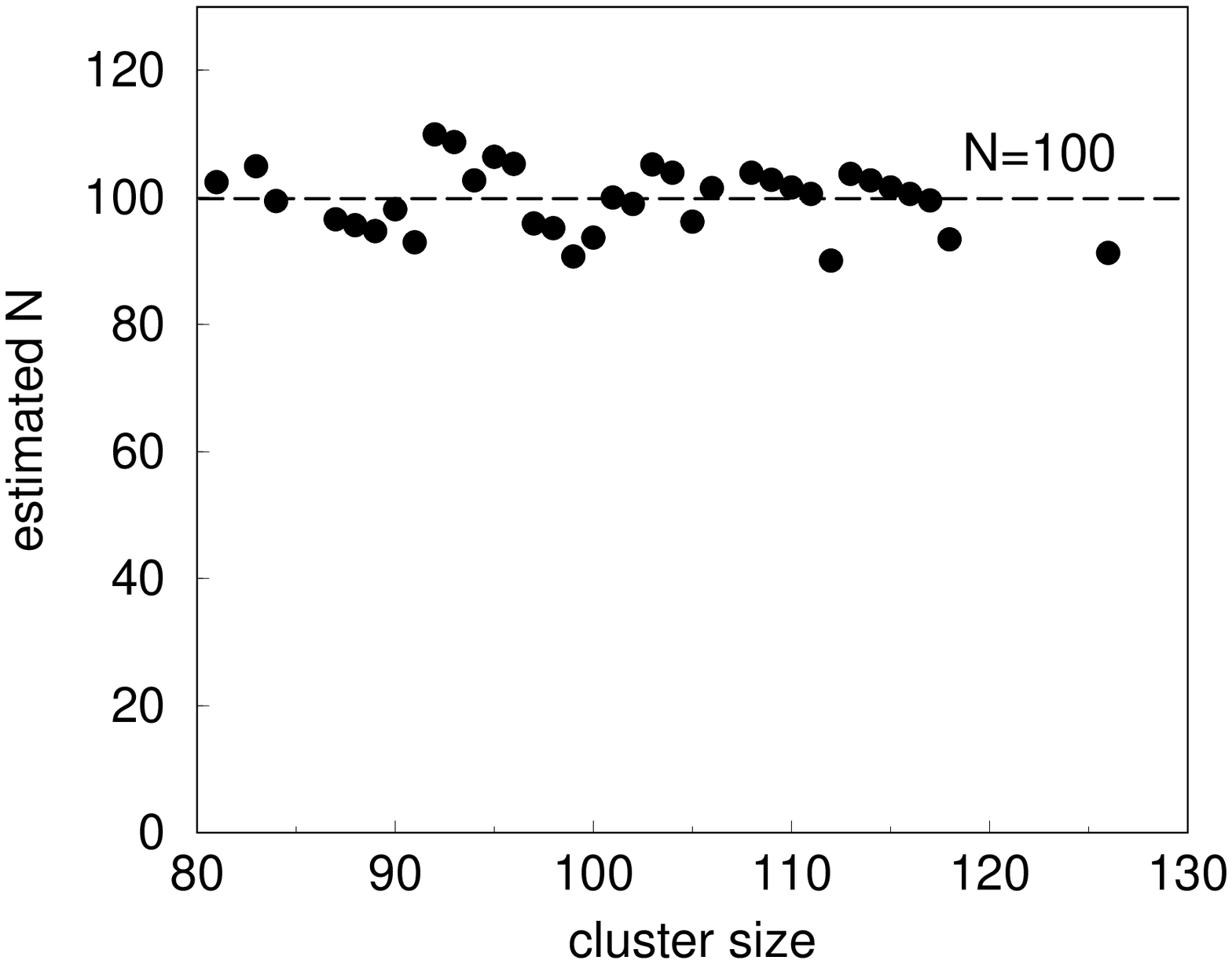}
\includegraphics{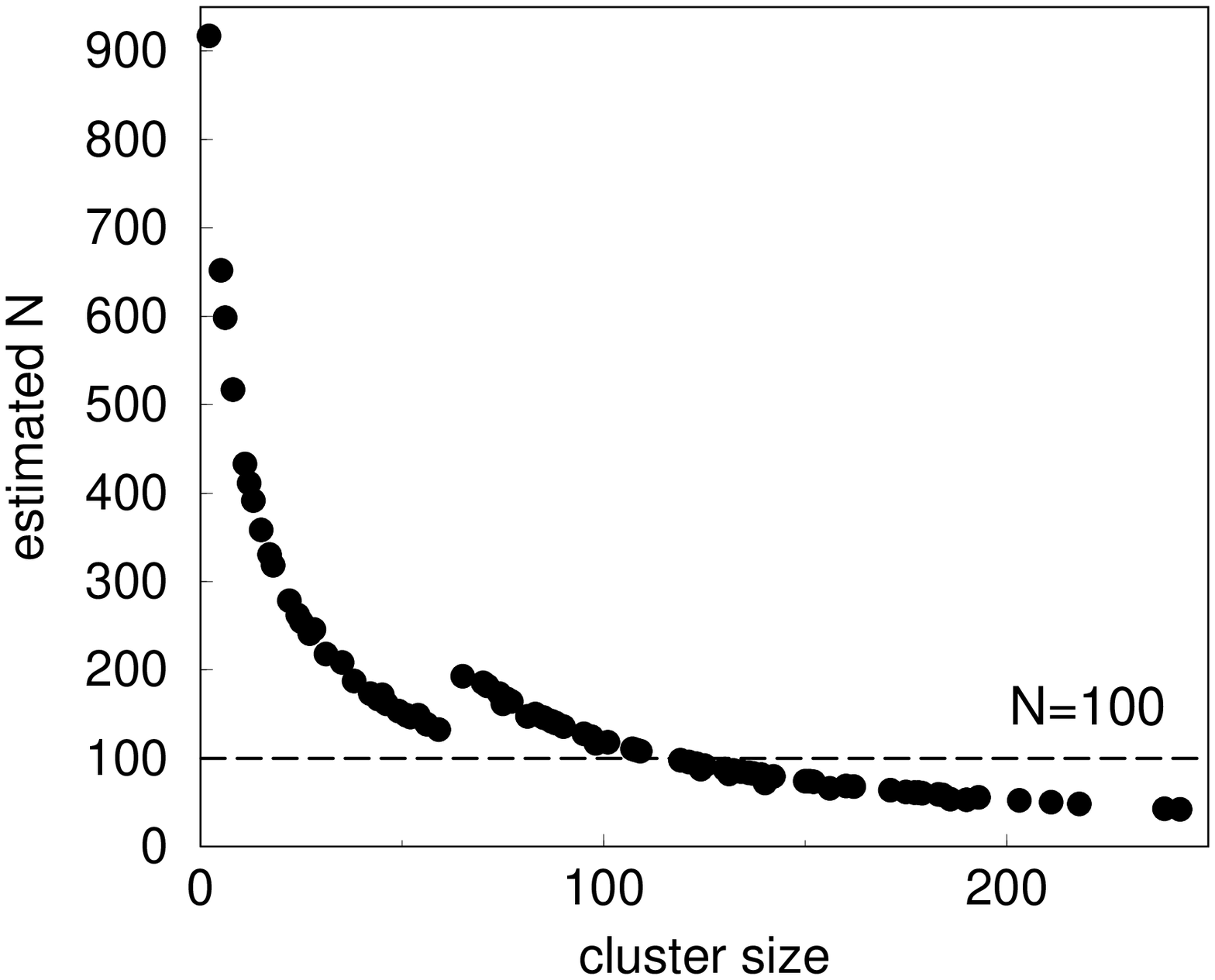}
  }   
  \resizebox{\columnwidth}{!}{
\includegraphics{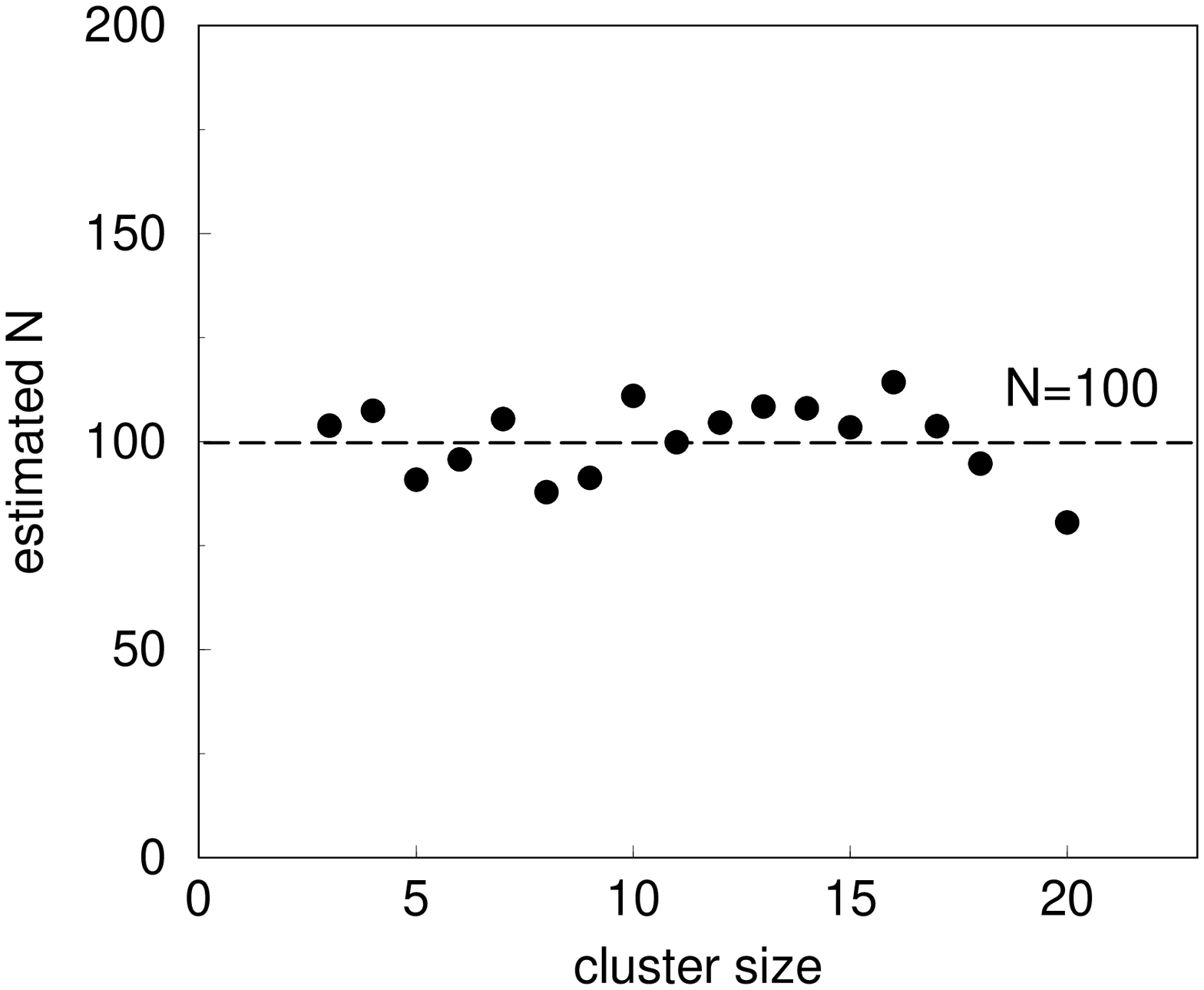}
\includegraphics{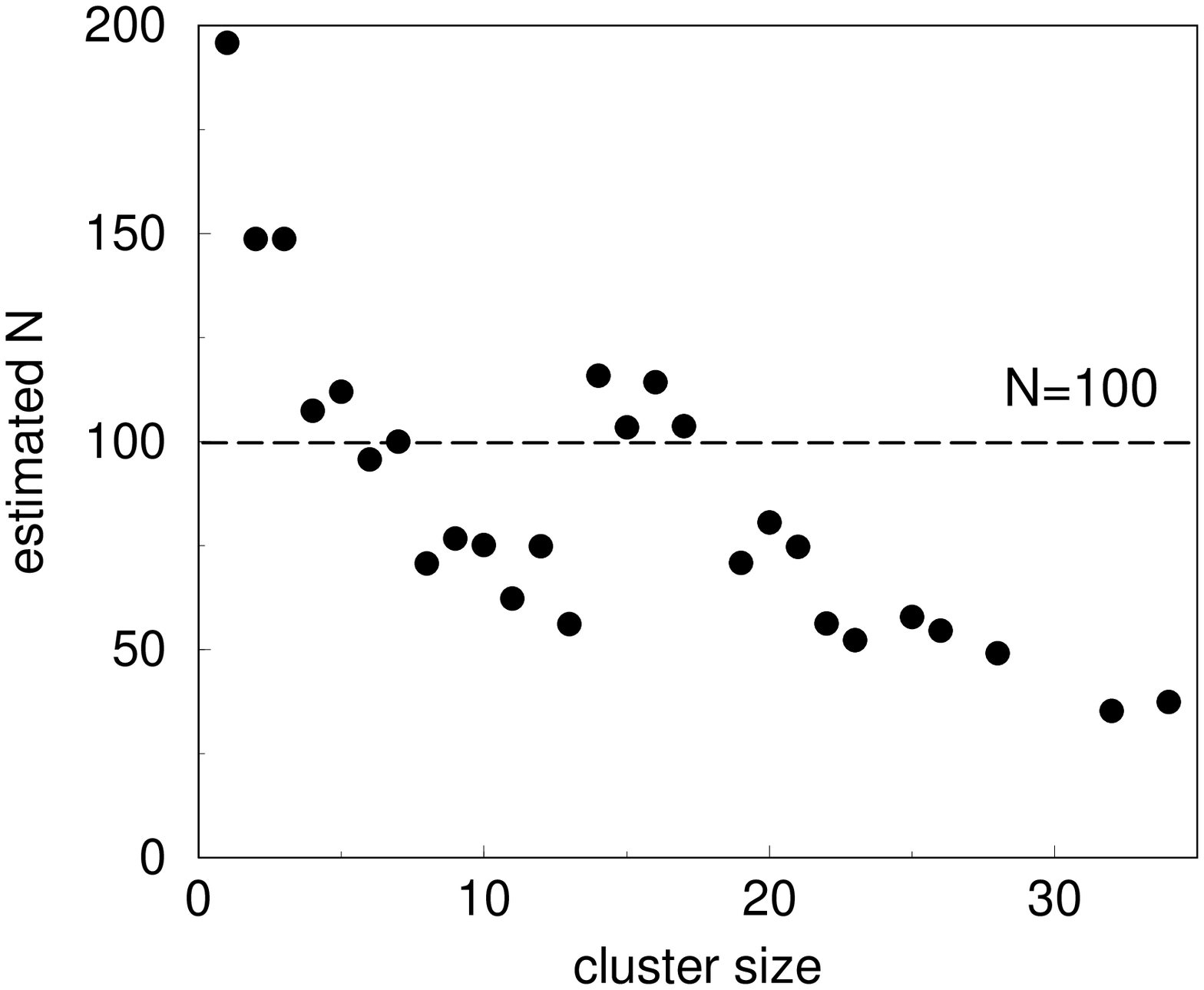}
  }   
  \resizebox{\columnwidth}{!}{
\includegraphics{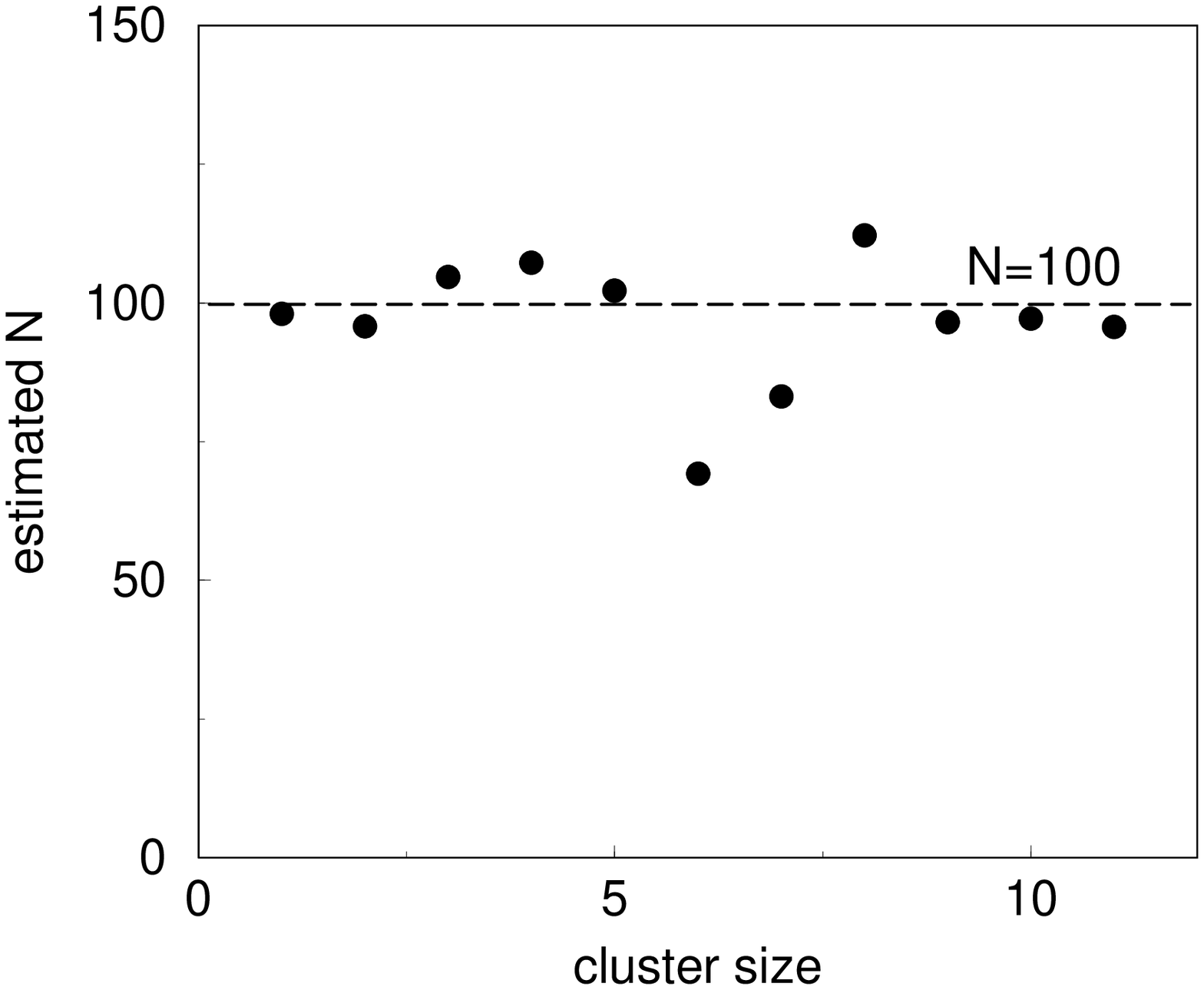}
\includegraphics{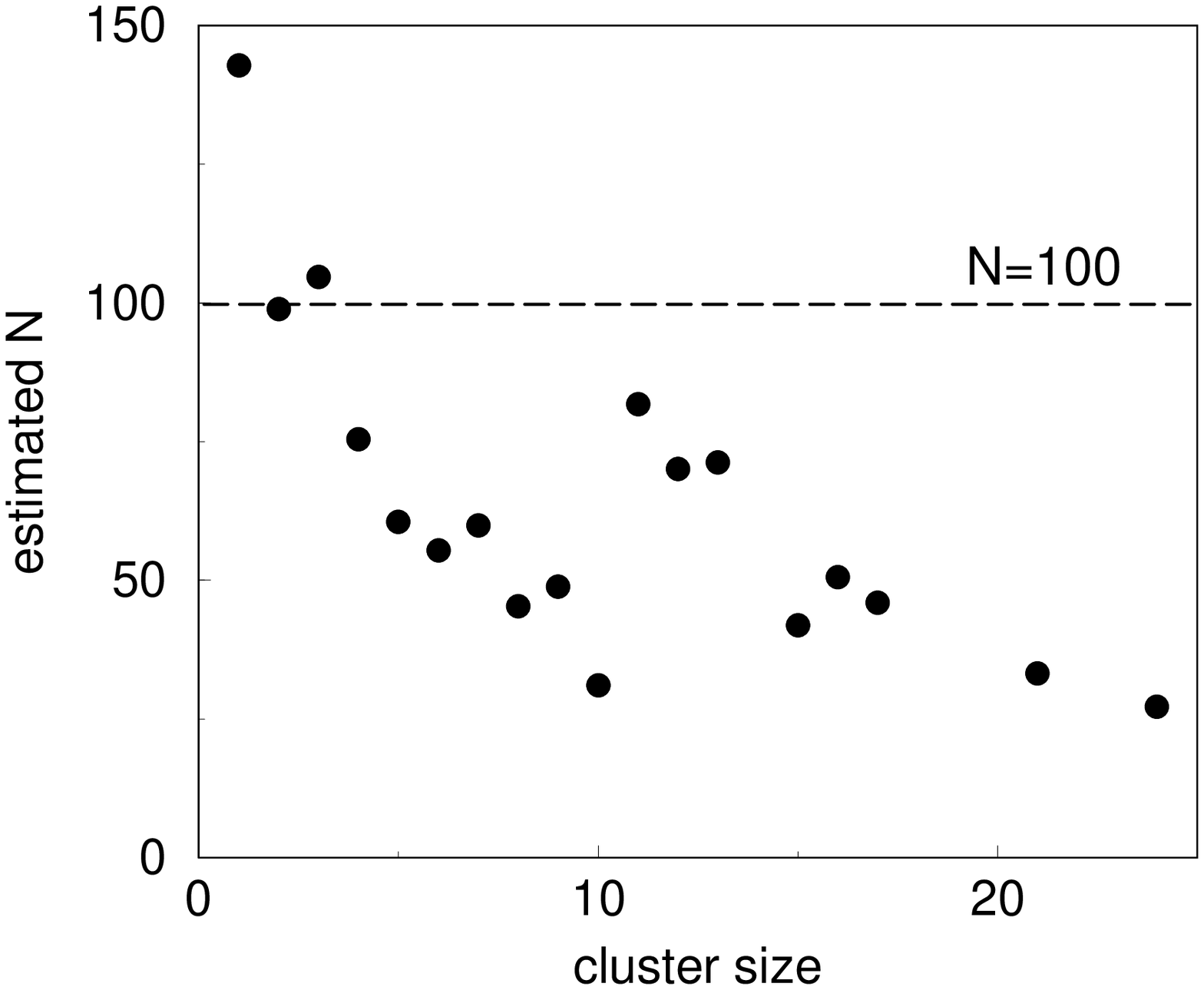}
  }   
  \resizebox{\columnwidth}{!}{
\includegraphics{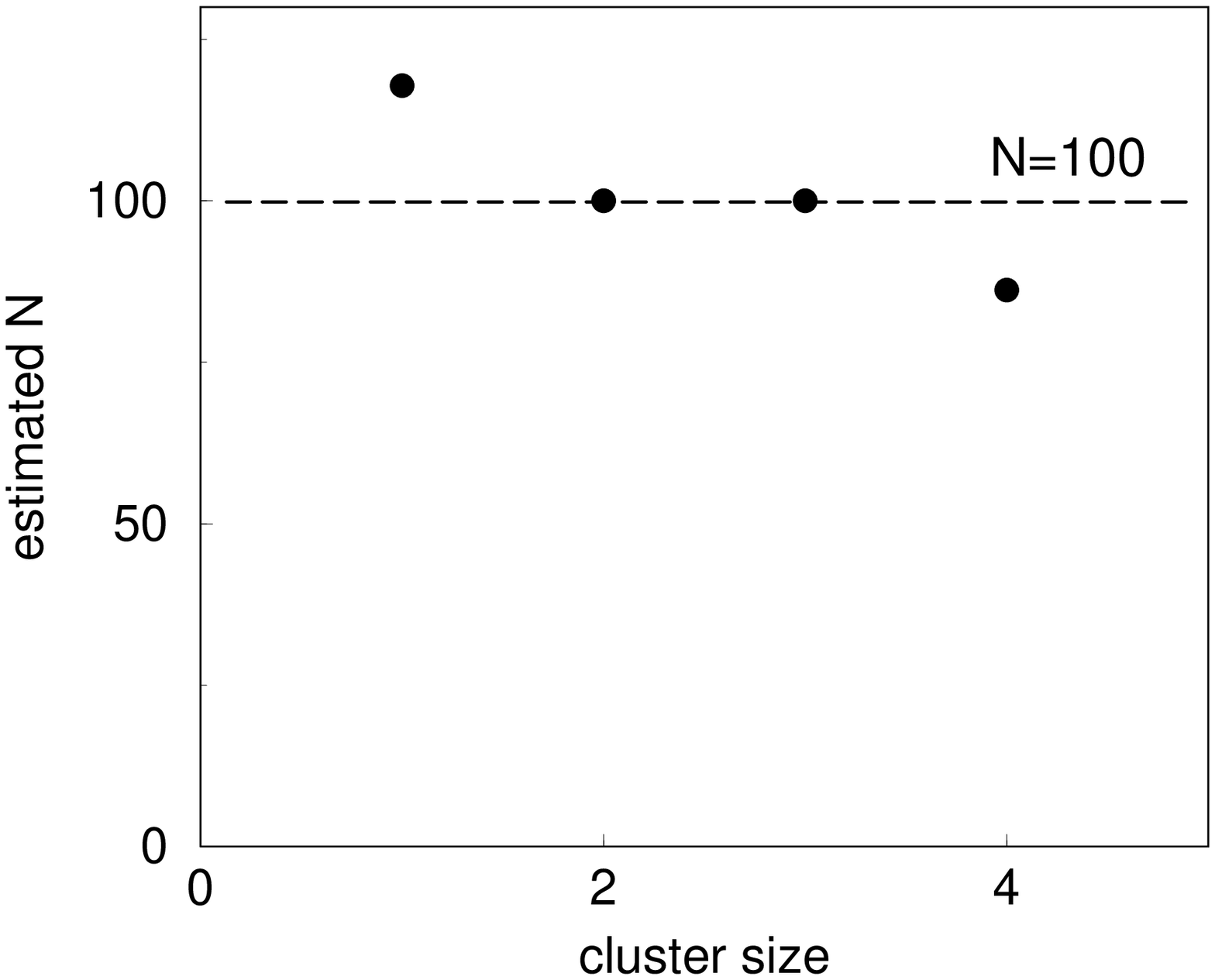}
\includegraphics{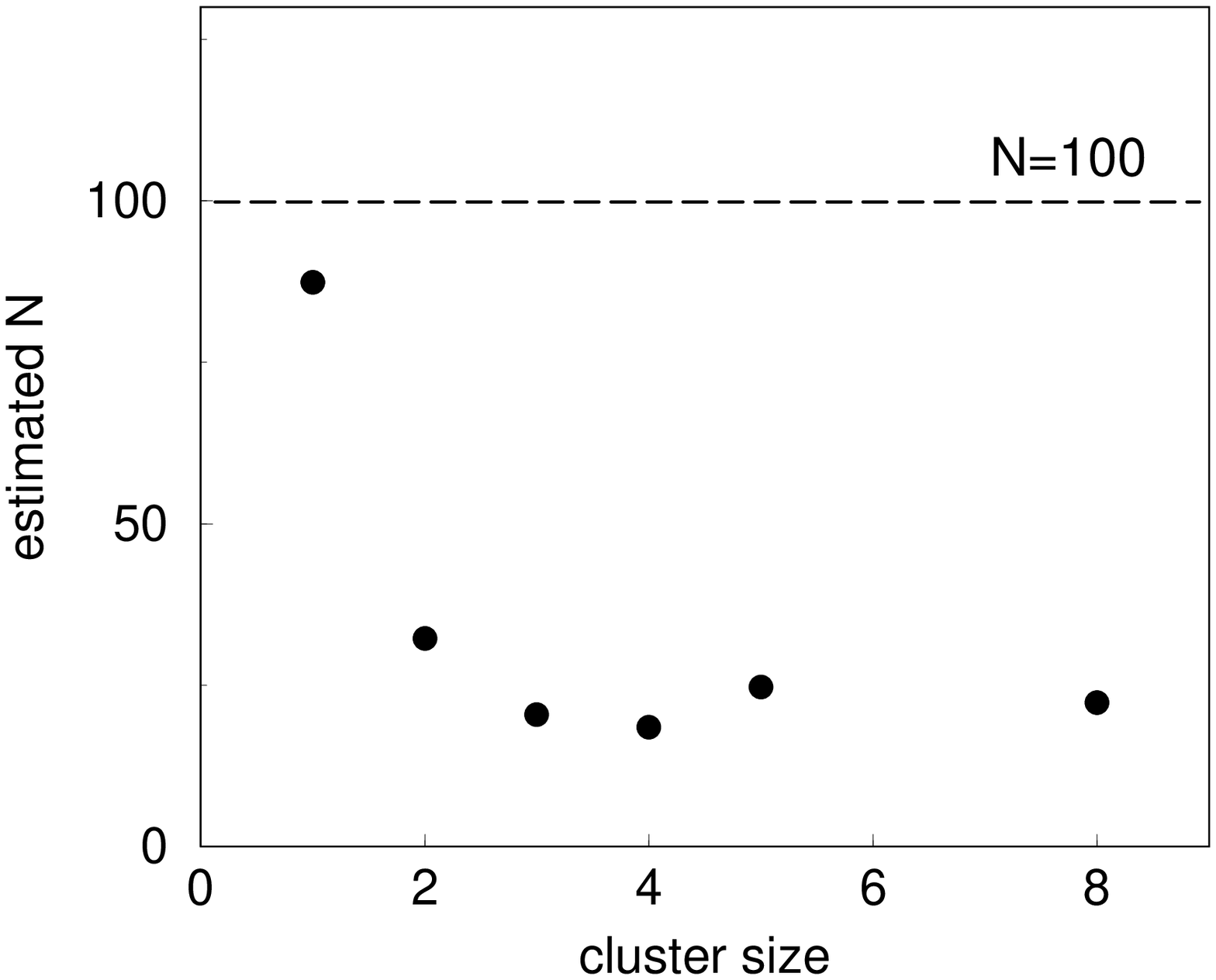}
  }   
\caption{Estimated total number of species due to Eqs. (\ref{eq:N1})
and (\ref{eq:Ni}) for the data shown in Fig. \ref{fig:deform}. For
equidistributed species (left figures) we find the condition
Eq. (\ref{eq:N}) fulfilled, whereas for triangularly distributed species
different occupation numbers $k_i$ yield significantly different estimates
$N^{(i)}$. From top to bottom: $M=10^4$, $M=10^3$, $M=500$, $M=100$. The dashed lines display the correct value $N=100$.}
\label{fig:Nest}
\end{figure}

For all plots drawn in Fig. \ref{fig:deform} we calculated the
estimated total number of events $N^{(i)}$ from the observed
occupation numbers $k_i$ for diverse cluster sizes $i$. The result is
shown in Fig. \ref{fig:Nest}. As expected, only for the
equidistributed species (left plots) we find that the approximate
constancy of $N^{(i)}$ as expressed by Eq. (\ref{eq:N}) holds
true. Surprisingly, even for quite small sample size $M=100$, i.e.,
very poor statistics, we can clearly distinguish between the frequency
distributions originating from the equidistribution
Eq. (\ref{eq:pequi}) and those which originate from the triangular
distribution Eq. (\ref{eq:ptriang}).

\begin{figure}[htbp]
\centerline{  \resizebox{7cm}{!}{
\includegraphics{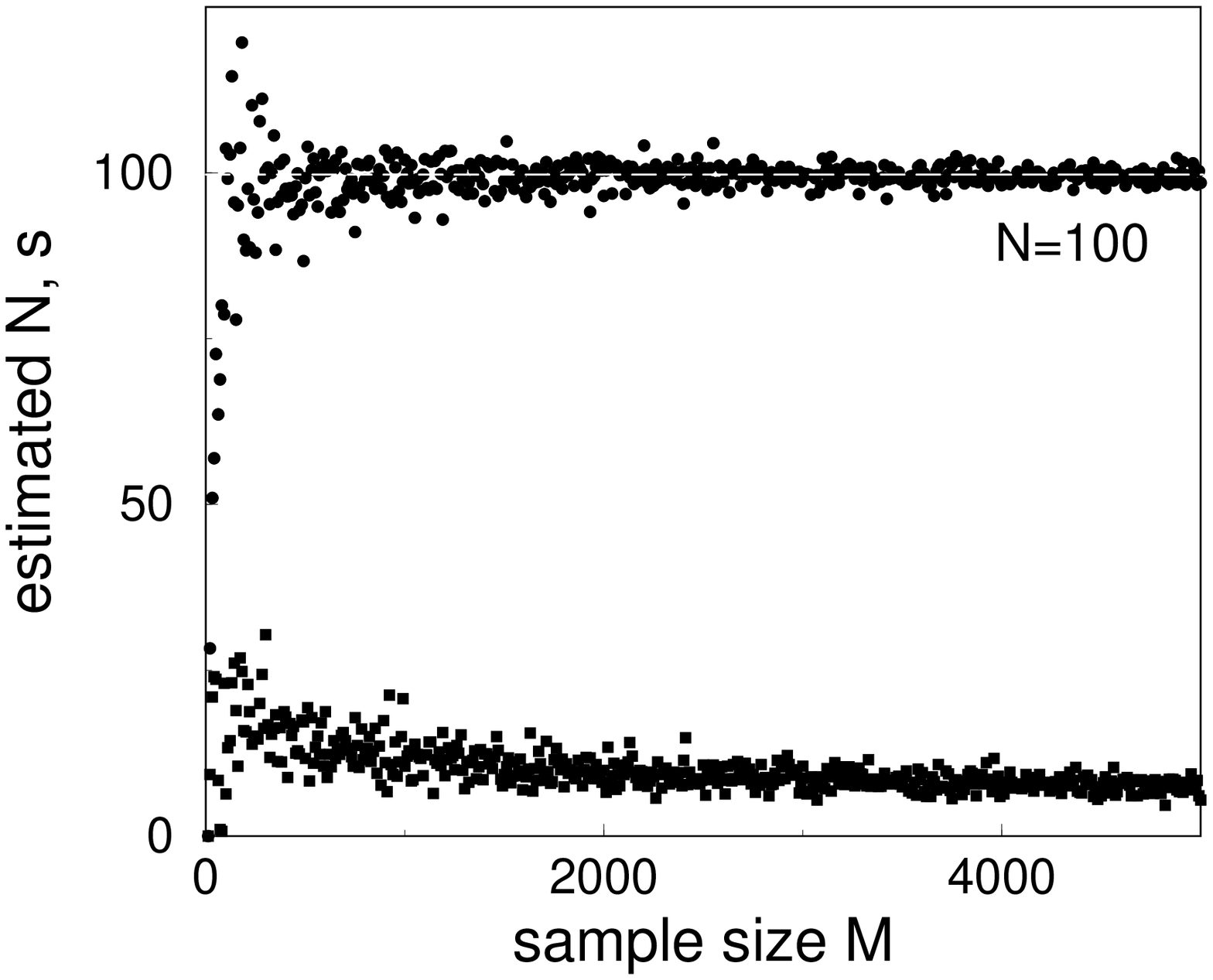}}
}
\centerline{  \resizebox{7cm}{!}{
\includegraphics{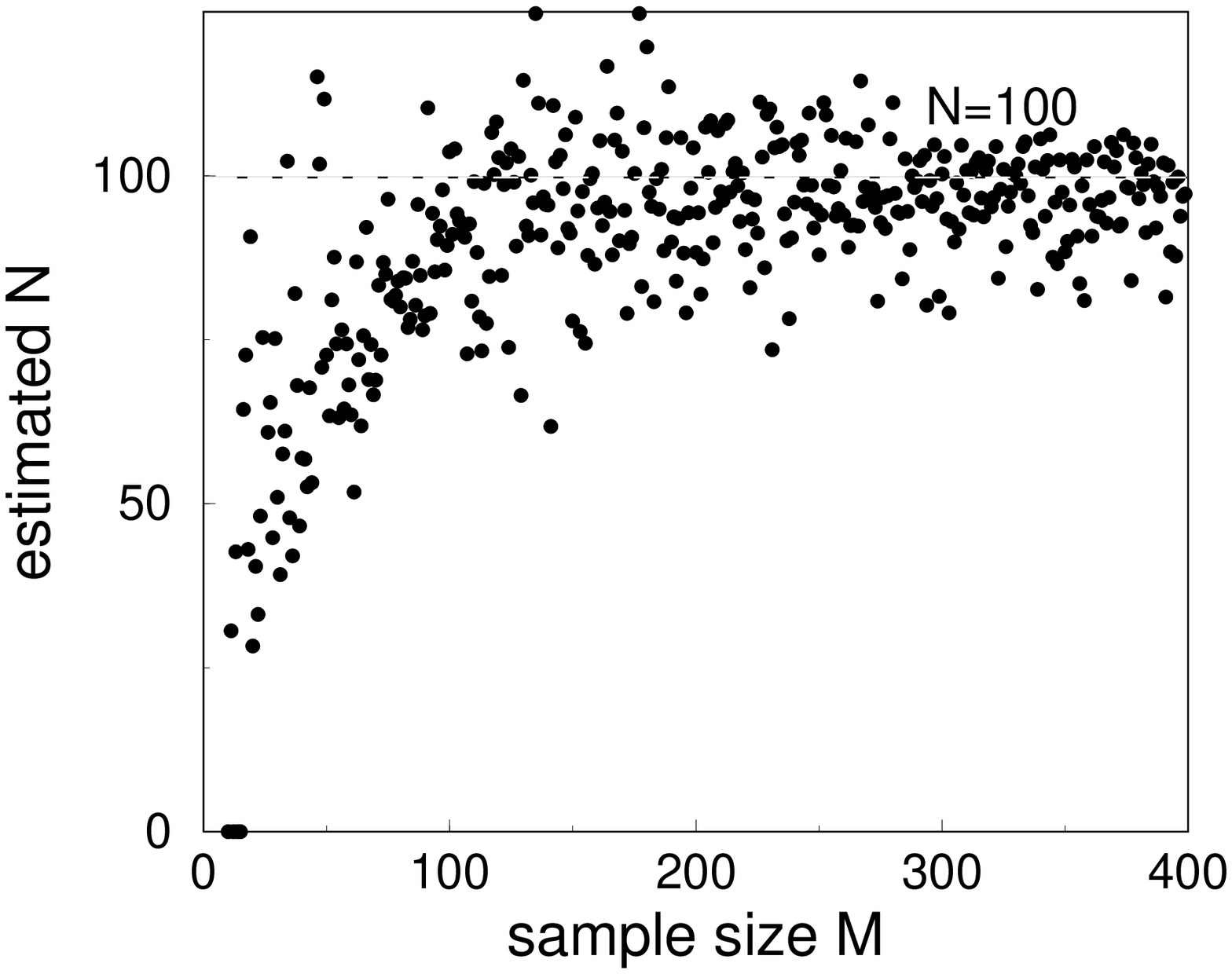}
  }   }
  \caption{The mean estimated total number of species (circles) and the standard deviation $s$ (boxes) due to Eqs. (\ref{eq:Nmean}) and (\ref{eq:StdDev}) for species distributed according to the equidistribution Eq. (\ref{eq:pequi}). The dashed lines show the true total number of species $N=100$.}
  \label{fig:EqNest}
\end{figure}

Figure \ref{fig:EqNest} shows the mean
\begin{equation}
  \label{eq:Nmean} N_{\rm est}\equiv \frac{\sum\limits_{i=1}^M
  N^{(i)}}{\#\left(k_i\right)}
\end{equation}
of the estimated total number of species and the standard deviation
\begin{equation}
  \label{eq:StdDev} s\equiv\sqrt{\frac{\sum\limits_{i=1}^M
  \left(N^{(i)}\right)^2}{\#\left(k_i\right)}-
  \left(\frac{\sum\limits_{i=1}^M
  N^{(i)}}{\#\left(k_i\right)}\right)^2}\,
\end{equation}
for the equidistribution over the sample size $M$. The summation in Eqs. (\ref{eq:Nmean}) and
(\ref{eq:StdDev}) runs over all indices $i\in [1, M]$ which correspond
to occupation numbers $k_i \ne 0$. The symbol
$\#\left(k_i\right)$ stands for the number of these
non-empty occupations. The larger the sample size the smaller the
standard deviation. Even for rather small sample sizes $M\approx 150$
the estimated total number of species agrees well with the true value
$N=100$ as shown in the zoomed region (lower plot in
Fig. \ref{fig:EqNest}). The dashed lines show the true total number of
species, $N=100$.

\begin{figure}[htbp]
\centerline{  \resizebox{7cm}{!}{
\includegraphics{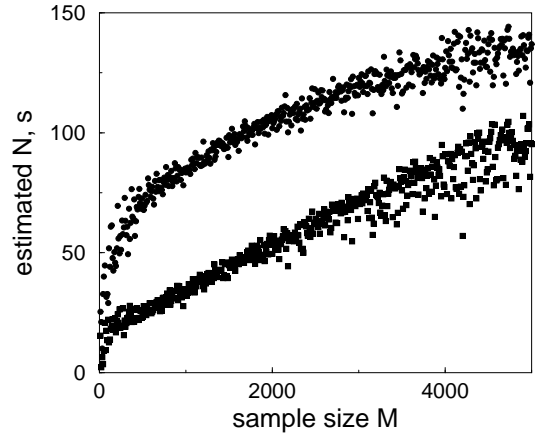}
  }   }
  \caption{The mean estimated total number of species (circles) and the standard
  deviation $s$ (boxes) due to Eqs. (\ref{eq:Nmean}) and (\ref{eq:StdDev}) for
  species distributed according to the triangular distribution
  Eq. (\ref{eq:ptriang}).}
\label{fig:DrNest}
\end{figure}

Figure \ref{fig:DrNest} depicts the corresponding data for species
which were generated from the triangular probability distribution
Eq. (\ref{eq:ptriang}). For larger sample sizes $M$ we obtain a
growing mean of the estimated total number of species. With
increasing sample size the standard deviation increases too.

\section{Solutions of Eq. (7)}
\label{sec.solutions}

\begin{figure}[htbp]
\centerline{  \resizebox{7cm}{!}{
\includegraphics[angle=270]{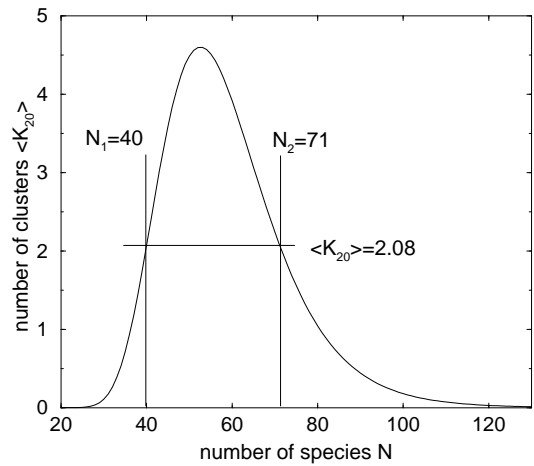}
  }   }
  \caption{$\left<K_{20}\right>$ over the total number of species $N$
  for $M=1000$. The curve has a maximum $\left<K_{20}\right>^{\rm
  max}\approx 4.6$ at $N\approx 52.6$.}  
  \label{fig:twosol1}
\end{figure}

From Eq. (\ref{eq:Mom}) it follows that for a given sample size $M$
the curve $\left<K_i\right>$, now understood as a function of the
total number of species $N$, has a maximum. As an example,
in Fig. \ref{fig:twosol1} we plotted $\left<K_{20}\right>$ vs.~$N$ (for
$M=1000$). The curve has a maximal value $\left<K_{20}\right>^{\rm
max}\approx 4.6$ at $N\approx 52.6$, meaning that there is no uniform
probability distribution which is in agreement with any larger
experimentally determined values of $k_{20}$.

\begin{figure}[htbp]
\centerline{  \resizebox{7cm}{!}{
\includegraphics[angle=270]{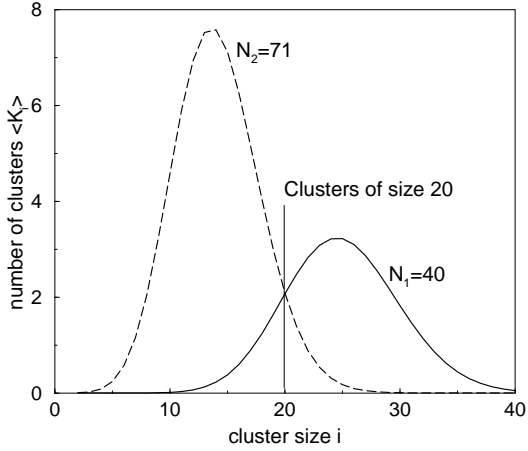}
  }   }
  \caption{$\left<K_i\right>$ over $i$ due to Eq. (\ref{eq:Mom}) for
  $M=1000$ and $N_1=40$ (full line) and $N=71$ (dashed line),
  respectively. The lines intersect at $i=20$.}
  \label{fig:twosol2}
\end{figure}

The extremum of $\left<K_i\right>$ for $i\ge 2$ can be found from
Eq. (\ref{eq:Mom}):
\begin{equation}
  \label{eq:Kextr}
\left<K_i\right>^{\rm max} = \left(M \atop i\right) (1-M)^{1-M} (i-M)^{M-i} (1-i)^{i-1}  
\end{equation}
which occurs for
\begin{equation}
  \label{eq:Nextr}
  N^{\rm max} = \frac{M-1}{i-1}\,.
\end{equation}
Hence, Eq. (\ref{eq:Ni}) has no solution for any $k_i >\left<K_i\right>^{\rm max}$. Figure \ref{fig:twosol1} shows also that
for any $k_i < \left<K_i\right>^{\rm max}$ there are two
solutions of Eq. (\ref{eq:Ni}): for $M=1000$ the value
$k_{20}=2$ has the solutions near $N_1= 40$ and
$N_2=71$. The reason for this behaviour becomes clear when drawing
$\left<K_{i}\right>$ over $i$ for $N_1=40$ and $N_2=71$
(Fig. \ref{fig:twosol2}). The curves intersect at $i=20$ in agreement
with the two solutions of Eq. (\ref{eq:Ni}). 
In the same way each value of $k_i <\left<K_i\right>^{\rm max}$ corresponds to two such solutions, $N_1^{(i)} \le N_2^{(i)}$. Hence,
for a construction of the plots in Fig.~\ref{fig:Nest} one always has
to choose the appropriate branch. If we had only one measurement $k_i$ for a particular $i$ there were no means to figure out whether the corresponding  $N$ is $N_1$ or $N_2$. In our case, however, there is a series of $k_i$ for a certain range of $i$ and all of them originate from the same measurement.

Obviously, the size $k_i$ of the largest $i$ which is found in the data corresponds to the right branch of the curve $\left<K_i\right>(i)$ as drawn in Fig. \ref{fig:twosol2}, i.e., to $N_2^{(i)}$. If it would correspond to the left branch, larger $i$ should still be populated. Contrary, the smallest $i$ correspond to $N_1^{(i)}$. From the single humped-shape of the curve $\left<K_i\right>(i)$ we conclude that there is only one transition from $N_1$ to $N_2$, i.e., the solution is
\begin{equation}
  \label{eq:select}
  N^{(i)}=\left\{
    \begin{tabular}{lll}
$N_2^{(i)}$ & for &$ i\le i^*$\\
$N_1^{(i)}$ & for &$ i>i^*$\,,\\ 
    \end{tabular}\right.
\end{equation}
where $i^*$ has to be determined. For an equidistribution we expect that $N^{(i)}$ for all $i$ approximate the true $N$ (see Eq. (\ref{eq:N})). Therefore, we choose $i^*$ from $\left[i_{\rm min},i_{\rm max}\right]$ such that it minimises
\begin{equation}
\left(\Delta N\right)^{\left(i^*\right)}\equiv \sqrt{\overline{\left(N^{\left(i^*\right)}\right)^2}-\left(\overline{N^{\left(i^*\right)}}\right)^2}\rightarrow \min\,,
\label{eq:min}
\end{equation}
with
\begin{eqnarray}
  \overline{N^{\left(i^*\right)}}&\equiv& \frac{\sum\limits_{i=i_{\rm min}}^{i^*} N_2^{(i)}+\sum\limits_{i=i^*+1}^{i_{\rm max}} N_1^{(i)}}{i_{\rm max}-i_{\rm min}+1} \label{Nav}\\
\overline{\left(N^{ \left(i^*\right)}\right)^2} & \equiv& \frac{\sum\limits_{i=i_{\rm min}}^{i^*} \left(N_2^{(i)}\right)^2+\sum\limits_{i=i^*+1}^{i_{\rm max}} \left(N_1^{(i)}\right)^2}{i_{\rm max}-i_{\rm min}+1}
\end{eqnarray}
and with $i_{\rm min}$ and $i_{\rm max}$ being the sizes of the smallest and largest clusters which are found in the data set.

\begin{figure}[htbp]
\centerline{  \resizebox{5cm}{!}{
\includegraphics{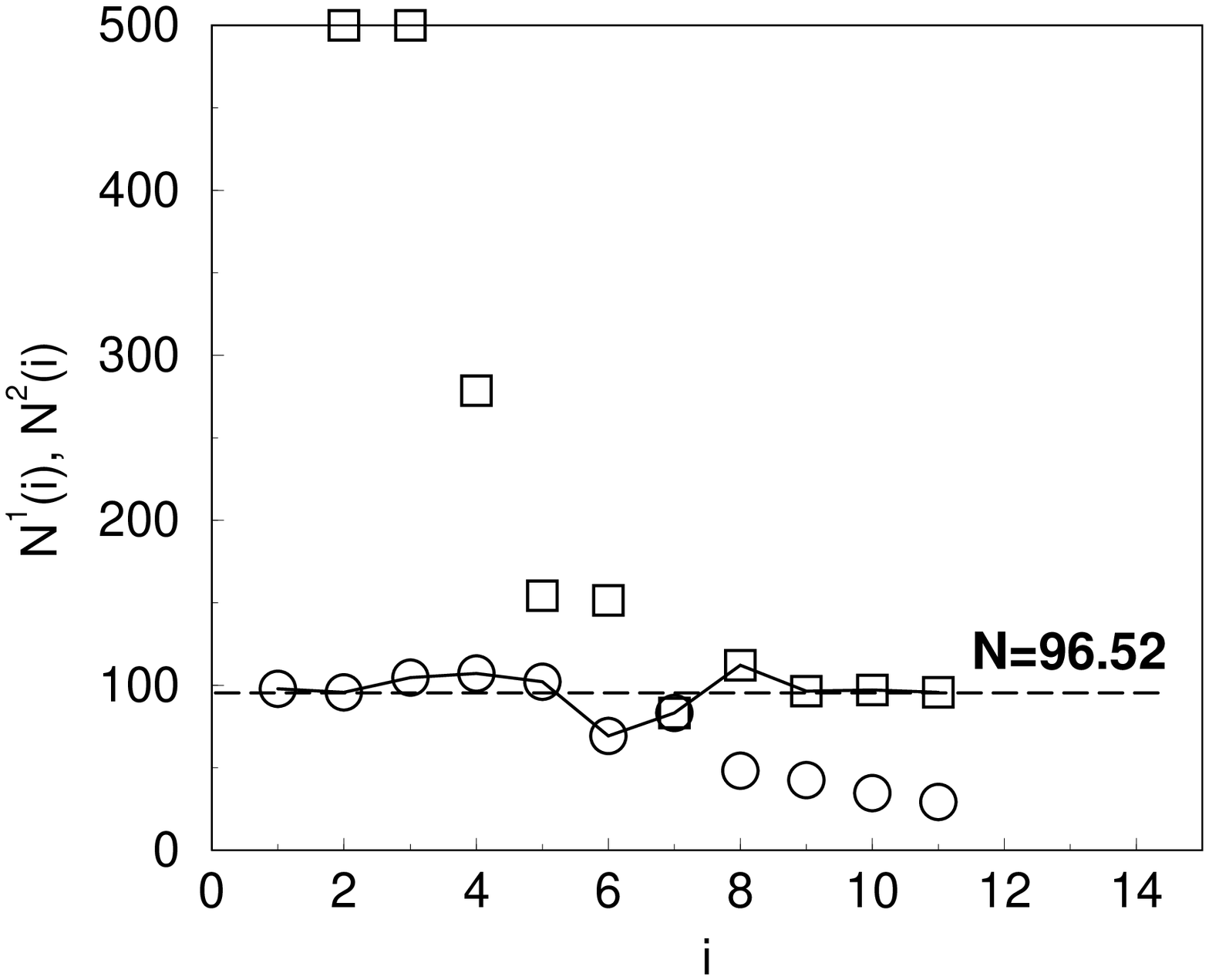}  
}  }
\centerline{\resizebox{5cm}{!}{
\includegraphics{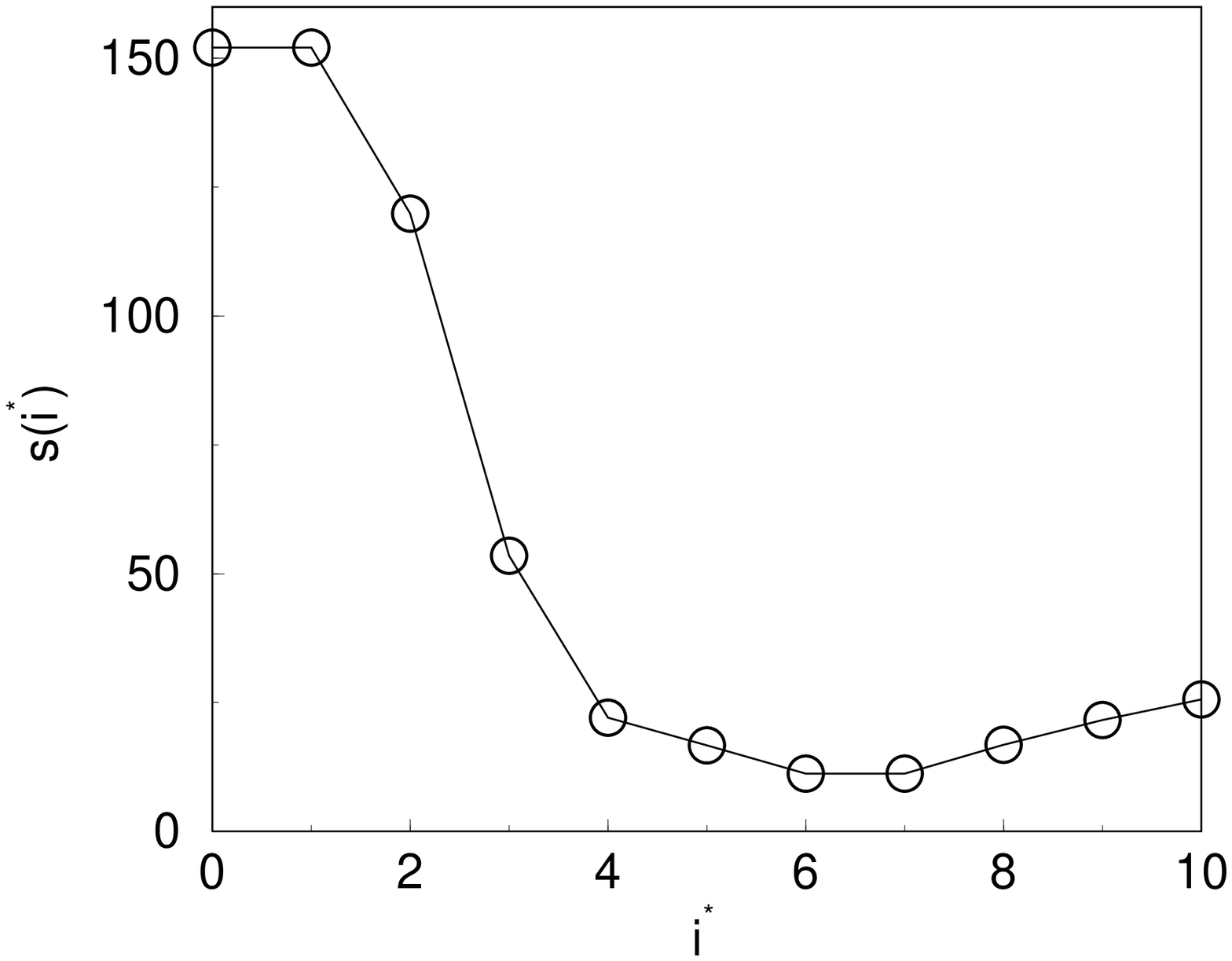}  
}}
  \caption{Top: Both solutions of Eq. (\ref{eq:Ni}), $N^1(i)$ and $N^2(i)$ for the equidistribution $M=500$ and $N=100$, corresponding to Fig. (\ref{fig:Nest}) (3rd row, left column). The choice $i^*=7$ minimises $s$ (see bottom figure) and yields $\overline{N}\approx 96.52$. }
  \label{fig:select}
\end{figure}

The estimated number $N$ which is drawn in Figs. \ref{fig:Nest}, \ref{fig:EqNest}, and \ref{fig:DrNest} is, hence, given by Eq. (\ref{eq:select}) with the condition (\ref{eq:min}). Figure (\ref{fig:select}) drafts the described procedure.

\section{Discussion}
Given a sampling probe of size $M$ originating from an unknown
probability distribution, it may be important to decide whether the
data set is compatible with an equidistribution with known or unknown
total number of species $N$. Due to finite sample size effects,
experimentally accessible frequency distributions will always
experience deviations from the underlying probability distribution -
be it uniform or non-uniform. Rank-ordering the frequency
distribution, being a first step towards a systematic survey, still
does not help for rather small sample sizes. The challenge is to find
a criterion, solely based on the frequency distribution (data set),
which allows to accept or reject the hypothesis of a uniform
probability distribution and, in case of acceptance, to render the
number $N$ within statistical errors.

We developed a method based on an analytic expression for the average
number of events $\left<K_i\right>$ occurring $i$ times (the varying
lengths of the plateaus in the rank-ordered frequency distributions)
which involves the sample size $M$ and the a priori (un)known number
$N$. By inversion of this relation it is possible to compute for each
$i$ an estimate $N^{(i)}$ for the hypothetical $N$, completely
specifying the assumed equidistribution. For true uniform probability
distributions these estimates should vary slightly, i.e.~within
expected statistical errors, when varying $i$. Substantial variations
of the estimates, i.e. beyond expected statistical errors, or trends
are a clear signature of a non-uniform distribution.

We exemplified this method by applying it to a uniform and a
triangular distribution. The separation between both by virtue of our
criterion is possible down to small sample sizes for which visual
inspection of rank-ordered frequency distributions or standard tests such as $\chi^2$ and Kolmogorov-Smirnov do not allow for a clear-cut distinction.

\acknote ~ 
We thank Werner Ebeling and Cornelius Fr\"ommel for discussion.


\end{document}